\pdfoutput=1
\documentclass[12pt]{article}

\usepackage[T1]{fontenc}
\usepackage{mathtools}
\usepackage{amssymb}
\usepackage{slashed}
\usepackage{multirow}
\usepackage{fullpage}
\usepackage[bottom]{footmisc}
\usepackage{longtable}
\usepackage{mathrsfs}
\usepackage{comment}
\usepackage{tipa}
\usepackage{tipx}
\usepackage{wasysym}

\usepackage{epsfig,latexsym,amsfonts,amsmath,amsthm,amssymb,amsbsy,multirow,slashed,wasysym,textcomp,cite,comment,subfigure,wrapfig,datetime,mathtools,cancel,mathrsfs,pict2e}
\usepackage{datetime2}
\usepackage[font={footnotesize,sl},bf]{caption}

\usepackage{mathtools}
\usepackage{mathrsfs}
\usepackage{setspace}
\usepackage{tikz}
\usetikzlibrary{calc}
\usetikzlibrary{decorations.pathreplacing}
\usetikzlibrary{decorations.pathmorphing}
\usetikzlibrary{decorations.markings}
\usetikzlibrary{arrows.meta}
\usetikzlibrary{positioning}

\usepackage{xcolor}
\definecolor{oldmauve}{rgb}{0.4, 0.19, 0.28}
\definecolor{pansypurple}{rgb}{0.47, 0.09, 0.29}
\definecolor{burgundy}{rgb}{0.5, 0.0, 0.13}
\definecolor{carminepink}{rgb}{0.92, 0.3, 0.26}
\definecolor{blue(pigment)}{rgb}{0.2, 0.2, 0.6}
\definecolor{darkseagreen}{rgb}{0.56, 0.74, 0.56}
\definecolor{darkspringgreen}{rgb}{0.09, 0.45, 0.27}
\definecolor{ceruleanblue}{rgb}{0.16, 0.32, 0.75}

\definecolor{navyblue}{RGB}{0,0,128}

\DeclareMathOperator{\extdm}{d}
\newcommand{\extd}{\extdm \!}

\numberwithin{equation}{section}

\def\bea{\begin{eqnarray}}
\def\eea{\end{eqnarray}}
\newcommand{\beq}{\begin{eqnarray}}
\newcommand{\eqq}{\end{eqnarray}}
 \newcommand{\badat}{\begin{alignedat}}
 \newcommand{\eadat}{\end{alignedat}}

\newcommand{\eal}[1]{\be \begin{aligned} #1 \end{aligned}\end{equation}} 
\newcommand{\eqn}[1]{\be #1 \end{equation}} 
\newcommand{\eqa}[1]{\bea  #1\end{eqnarray}}

\newcommand{\sq}[1]{[#1]}

\newcommand{\cO}{{\cal O}}

\long\def\new#1\endnew{{\bf #1}}		
\long\def\del#1\enddel{}

\def\del{\partial}

\def\ndelta{\delta\hspace{-0.50em}\slash\hspace{-0.05em} }

\newcommand{\be}{\begin{eqnarray}}
\newcommand{\en}{\end{eqnarray}}

\newenvironment{align1}
{\begin{equation}\begin{aligned}}
{\end{aligned}\end{equation}}

\def\sq{\sqrt{q}}

\def\ndelta{\delta\hspace{-0.50em}\slash\hspace{-0.05em} }

\numberwithin{equation}{section} % equation numbers follow sections

\usepackage{pict2e}
\makeatletter
\newcommand{\loplus}{\mathbin{\mathpalette\dog@lsemi{+}}}
\newcommand{\dog@lsemi}[2]{\dog@semi{#1}{#2}{270,90}}
\newcommand{\dog@semi}[3]{%
  \begingroup
  \sbox\z@{$\m@th#1#2$}%
  \setlength{\unitlength}{\dimexpr\ht\z@+\dp\z@\relax}%
  \makebox[\wd\z@]{\raisebox{-\dp\z@}{%
    \begin{picture}(1,1)
    \linethickness{\variable@rule{#1}}
    \roundcap
    \put(0.5,0.5){\makebox(0,0){\raisebox{\dp\z@}{$\m@th#1#2$}}}
    \put(0.5,0.5){\arc[#3]{0.5}}
    \end{picture}%
  }}%
  \endgroup
}
\newcommand{\variable@rule}[1]{%
  \fontdimen8  
  \ifx#1\displaystyle\textfont3\else
    \ifx#1\textstyle\textfont3\else
      \ifx#1\scriptstyle\scriptfont3\else
        \scriptscriptfont3\relax
  \fi\fi\fi
}
\makeatother

\newcommand{\pthorn}{{\text{\th}}}

\author{}
\numberwithin{equation}{section} % equation numbers follow sections
\usepackage{hyperref}
\usepackage{orcidlink}

\begin{document}

\begin{titlepage}

  \thispagestyle{empty}

 \begin{flushright}
%NORDITA 2022-084 \\
% UUITP-55/22 \\
 \end{flushright}

%\vskip2cm

  \begin{center}

%{\Large\textbf{Celestial $Lw_{1+\infty}$ Symmetries and Subleading Phase Space
%of Null Hypersurfaces}}

{\LARGE\textbf{Celestial $Lw_{1+\infty}$ Symmetries and Subleading}}
\vskip0.2cm
{\LARGE\textbf{Phase Space
of Null Hypersurfaces}}

\vskip0.8cm 
Romain Ruzziconi\footnote{\fontsize{8pt}{10pt}\selectfont\ \href{mailto:romainruzziconi@fas.harvard.edu}{romainruzziconi@fas.harvard.edu}} ,
C\'eline Zwikel\footnote{\fontsize{8pt}{10pt}\selectfont\ \href{mailto:celine.zwikel@college-de-france.fr}{celine.zwikel@college-de-france.fr}}
\vskip0.1cm

\normalsize
\medskip

%$^{1}$\textit{Mathematical Institute, University of Oxford, \\ Andrew Wiles Building, Radcliffe Observatory Quarter, \\
%Woodstock Road, Oxford, OX2 6GG, UK \\ \vspace{2mm}}
{\footnotesize{
$^{1}$\textit{Center for the Fundamental Laws of Nature, Harvard University \\
17 Oxford Street, Cambridge, MA 02138, USA} \\

\vspace{2mm}

$^{1}$\textit{Black Hole Initiative, Harvard University \\
20 Garden Street, Cambridge, MA 02138, USA}

\vspace{2mm}

$^2$\textit{Perimeter Institute for Theoretical Physics,\\ 31 Caroline Street North, Waterloo, Ontario, Canada N2L 2Y5}\\

\vspace{2mm}
$^2$\textit{Université Libre de Bruxelles and International Solvay Institutes,\\
ULB-Campus Plaine CP231,
B-1050 Brussels, Belgium}
\\
\vspace{2mm}
$^2$ \textit{Coll\`ege de France, 11 place Marcelin Berthelot, 75005 Paris, France}\\
}}
\end{center}

\vskip0.5cm

\begin{abstract}
\noindent Pursuing our analysis of \cite{Ruzziconi:2025fct}, we study the gravitational solution space around a null hypersurface in the bulk of spacetime, such as a black hole or a cosmological horizon. We discuss the corresponding characteristic initial value problem both in the metric and Newman-Penrose formalisms, and establish an explicit dictionary between the two. This allows us to identify Weyl-covariant structures in the solution space, including hierarchies of recursion relations encoding the flux-balance laws. We then establish a correspondence between the gravitational phase space at null infinity and the subleading phase space around the null hypersurface at finite distance. This connection is naturally formulated within the Newman-Penrose formalism by performing a partially off-shell conformal compactification and identifying the analogue of the Ashtekar-Streubel symplectic structure in the radial expansion near the null hypersurface. Using this framework, we identify the celestial $Lw_{1+\infty}$ symmetries in the subleading phase space at finite distance by constructing their canonical generators and imposing self-duality conditions. This allows us to define a notion of covariant radiation, whose absence gives rise to an infinite tower of conserved charges, revealing physical quantities relevant to observers near black hole or cosmological horizons. As a concrete illustration, we consider the case of the self-dual Taub–NUT black hole.

\end{abstract}

\end{titlepage}
\setcounter{page}{2}

\setcounter{tocdepth}{2}
\tableofcontents

\newpage
\section{Introduction}

The construction of the phase space of gravity in asymptotically flat spacetimes at null infinity and the identification of the associated symmetries has generated a lot of interests in the past decades. Key steps included the discovery of the BMS group as the asymptotic symmetry group \cite{Bondi:1962px,Sachs:1962zza,Sachs:1962wk}, the conformal compactification formalizing the notion of null infinity \cite{Penrose:1962ij, Penrose:1964ge,Newman:1966ub, Penrose:1986uia}, the complete characterization of the solution space and characteristic initial value problem \cite{Newman:1962cia,Tamburino:1966zz}, the construction of the radiative phase space and surface charges \cite{Ashtekar:1981bq,Wald:1999wa,Barnich:2011mi}, and the corresponding derivation for the various extensions of BMS \cite{Barnich:2010eb,Barnich:2011ct,Flanagan:2015pxa,Campiglia:2014yka,Campiglia:2015yka,Compere:2018ylh,Campiglia:2020qvc,Ruzziconi:2020cjt,Freidel:2021fxf,Geiller:2022vto,Fuentealba:2022xsz,Campoleoni:2023fug,Geiller:2024amx,Fiorucci:2024ndw,Geiller:2024ryw}. The understanding of symmetries and phase space at null infinity has been shown to be of major importance for the infrared sector of gravity \cite{Strominger:2017zoo} and the quest for flat space holography. Remarkably, the celestial $Lw_{1+\infty}$ symmetries \cite{Guevara:2021abz,Strominger:2021mtt}, appearing in Penrose's non-linear graviton construction of self-dual spacetimes from twistor space \cite{Penrose:1976jq,Penrose:1976js,Adamo:2021lrv}, have recently been included in this framework \cite{Freidel:2021dfs,Freidel:2021ytz,Campiglia:2021srh,Geiller:2024bgf,Kmec:2024nmu,Ruzziconi:2024kzo,Miller:2025wpq,Cresto:2024fhd,Cresto:2024mne,Kmec:2025ftx}, and their implications for scattering amplitudes and holography is still under investigation.  

By comparison, the phase space and symmetries of gravity on a null hypersurface at finite distance in the bulk, such as black hole or cosmological horizons, are much less understood. An important part of the complication arises from the fact that the intrinsic geometry of null hypersurfaces contains genuine degrees of freedom of the gravitational field, which must be included in the phase space. Their dynamics is described by the Raychaudhuri \cite{PhysRev.98.1123} and Damour \cite{PhysRevD.18.3598,Damour:1979wya} equations. This contrasts with null infinity, where these degrees of freedom are frozen by Einstein's equations and boundary conditions. Most works discussing physics on null hypersurfaces have focused on the intrinsic geometry, which we will refer to as the ``leading phase space''. This includes the characterization of the dynamics \cite{Ashtekar:1999yj,Ashtekar:2004cn,Ashtekar:2000hw,Ashtekar:2000sz,Booth:2012xm,Wieland:2017zkf}, the asymptotic symmetries \cite{Hawking:2016msc,Hawking:2016sgy,Carlip:2017xne,Haco:2018ske,Donnay:2015abr,Donnay:2016ejv,Chandrasekaran:2018aop,Grumiller:2019fmp,Adami:2020amw,Adami:2021nnf,Chandrasekaran:2021hxc,Odak:2022ndm,Odak:2023pga,Liu:2022uox}, the Carrollian geometry \cite{Mars:1993mj,Ciambelli:2019lap,Donnay:2019jiz,Redondo-Yuste:2022czg,Marsot:2022qkx,Freidel:2022bai,Freidel:2022vjq,Adami:2023wbe,Chandrasekaran:2023vzb,Freidel:2024emv}, and the phase space quantization \cite{Reisenberger:2007ku,Ciambelli:2023mir, Ciambelli:2024swv}. 

Interestingly, recent works \cite{Ashtekar:2024mme, Ashtekar:2024bpi, Riello:2024uvs, Ashtekar:2024stm} have suggested a direct connection between the phase space at null infinity and the leading phase space of a weakly isolated horizon, and have proposed a framework to treat both systems simultaneously. In the same vein, for scalar perturbations on the extreme Reissner-Nordstr\"om black hole, null infinity can be mapped onto the horizon using a conformal isometry generated by a spatial inversion \cite{Couch1984ConformalIU}. This can be used \cite{Lucietti:2012xr,Bizon:2012we} to map the Newman-Penrose conserved charges at null infinity \cite{Newman:1968uj} onto the Aretakis charges of the extremal horizon \cite{Aretakis:2011ha,Aretakis:2011hc,Aretakis:2012ei,Aretakis:2012bm}, and this connection was further extended to STU black holes in \cite{Godazgar:2017igz}, and to an infinite set of conserved quantities for
gravitational perturbations in \cite{Agrawal:2025fsv}. Furthermore, a treatment of null infinity as a stretched horizon was recently discussed in \cite{Riello:2024uvs} to renormalize the symplectic structure using Penrose's conformal compactification and provide a unified description of null hypersurfaces. Recent analyses relating the Brown--York Carrollian stress tensor at finite distance \cite{Chandrasekaran:2021hxc} and at null infinity \cite{Donnay:2022aba,Donnay:2022wvx} have appeared in \cite{Bhambure:2024ftz,Ciambelli:2025mex}.

In this work, pursuing our analysis initiated in \cite{Ruzziconi:2025fct}, we show that the Ashtekar-Streubel phase space at null infinity \cite{Ashtekar:1981bq} is directly related to the ``subleading radiative phase'' around a null hypersurface at finite distance. This allows us to import results from null infinity to the horizon,\footnote{Most of the discussion in this paper applies to generic null hypersurfaces, not only to black hole or cosmological horizons. For conciseness, we
will slightly abuse the terminology of “horizons” throughout the
text and use it for generic null hypersurfaces. Our end goal is to apply this general framework to actual black hole or cosmological horizons.} including the characterization of radiation, the recursion relations encoding the flux-balance laws, and the identification of the celestial \( Lw_{1+\infty} \) symmetries. This work opens new directions for studying the properties of black holes and cosmological horizons, and understanding Carrollian and celestial holography for finite regions. We summarize below our main results.

\paragraph{Summary of the results} $(i)$ First, we provide a complete characterization of the solution space at the horizon by discussing the characteristic initial value problem. We present these results in both metric and Newman-Penrose (NP) \cite{Newman:1961qr} formalisms and provide the explicit dictionary. This allows us to define a covariant notion of transverse radiation through the horizon, by analogy with null infinity. We also discuss the leading and subleading covariant phase space analysis in the metric formalism. $(ii)$ We then write the Bianchi identities in a Weyl-covariant form using the Geroch-Held-Penrose (GHP) operators \cite{Geroch:1973am,Penrose:1986uia}. This allows us to repackage the subleading flux-balance laws in a compact form and identify the analogue of the spin-$s$ recursion relations of \cite{Freidel:2021ytz, Geiller:2024bgf} at the horizon. $(iii)$ Exploiting the Weyl covariance of the GHP formalism, we perform a partially off-shell Penrose conformal compactification \cite{Penrose:1962ij, Penrose:1964ge,Newman:1966ub, Penrose:1986uia} and map the Peeling theorem and recursion relations at null infinity onto, respectively, the Taylor expansion of the Weyl tensor and recursion relations at the horizon. Furthermore, we show that, upon imposing self-duality conditions, the subleading symplectic structure in the radial expansion at the horizon is directly related to the Ashtekar-Streubel symplectic structure \cite{Ashtekar:1981bq} at null infinity. $(iv)$ We construct a tower of subleading spin-$s$ charges at a finite cut of the horizon and show that they are conserved in the self-dual subsector of gravity in the absence of transverse radiation. We then show that their associated integrated fluxes are indeed the canonical generators of $Lw_{1+\infty}$ symmetries at the horizon. Finally, we apply these general considerations to the case of a self-dual Taub–NUT black hole. 

\paragraph{Organization of the paper} The rest of the paper is organized as follows. In Section \ref{sec:Phase space in metric formalism}, we study the solution space of gravity in metric formalism around a generic null hypersurface and discuss the characteristic initial value problem. We also derive the subleading phase space near a null hypersurface by using the standard covariant phase space methods. In Section \ref{sec:From metric to Newman-Penrose formalism}, we construct an analogue of the Newman-Unti tetrad and translate the entire solution space into first-order Newman-Penrose formalism. We show that some of the evolution equations can be rewritten in a very compact and Weyl-covariant form in this formalism. In Section \ref{sec:From null infinity to the horizon}, we exploit the Weyl covariance of the formalism to match null infinity onto the horizon. Upon imposing self-duality conditions we identify the Ashtekar-Streubel symplectic structure in the subleading phase space at finite distance. We then construct the tower of $Lw_{1+\infty}$ charges and show that they generate the corresponding symmetries at the horizon. We show that the self-dual Kleinian Taub-NUT black hole is part of our phase space. In Section \ref{sec:Discussion}, we conclude with some discussions and possible implications of our work. The paper is also completed with several appendices: Appendix \ref{sec:Kerr} expresses the Kerr(-AdS) black hole solution around the horizon and provides a concrete example of the charges we are constructing in this paper. Appendix \ref{sec:Relevant equations} displays some relevant equations in the NP formalism. Appendix \ref{sec:Diffeomorphism interpretation} discusses the relation between the residual gauge diffeomorphisms at the horizon and the spin $s=0,1$ symmetries. Finally, Appendix \ref{sec:Radial expansion} discusses the identification of the higher-spin charge aspects in the radial expansion.

\section{Solution space in metric formalism}
\label{sec:Phase space in metric formalism}

In this section, we revisit the general solution space of general relativity around null hypersurfaces at finite distance, using the second-order metric formalism (see e.g. \cite{Hopfmuller:2016scf,Adami:2021nnf} for earlier works). This analysis allows us to fully characterize the data required to reconstruct the entire solution space around a null hypersurface. We also discuss leading and subleading phase space around a null hypersurface. We compute the residual symmetries at their associated charges.

\subsection{Gauge choice}

In Gaussian null coordinates $x^\mu = (v,r,x^A)$, $A = 1,2$, with gauge fixing conditions $g_{rr}=0=g_{rA}$ and $g_{vr}=1$, the line element reads as
\begin{equation}\label{metricGNC}
\extd s^2=-V\extd v^2+2\, \extd v\,dr+\gamma_{A B}\left(\extd x^A-U^A \extd v\right)\left(\extd x^B-U^B \extd v\right)\,.    
\end{equation} This is the analogue of the Newman-Unti gauge fixing usually adopted at null infinity \cite{Newman:1962cia,Barnich:2011ty}. Here, the bulk hypersurface of interest is at $r=0$, and we assume Taylor expansion of the transverse metric $\gamma_{AB}$ for $r>0$:
\begin{equation}\label{MetricTaylorexpansion}
    \gamma_{AB}=q_{AB}(v,x^A)+r\left(\chi_{AB}+\frac12 q_{AB} \chi\right)+\sum^{\infty}_{n=2} r^n\left(\chi^{(n)}_{AB}+\frac12 q_{AB} \chi^{(n)}\right)
\end{equation}
where $\chi_{AB}$ and $\chi_{AB}^{(n)}$ are trace-free tensors, i.e. $q^{AB}\chi_{AB} = 0 = q^{AB} \chi_{AB}^{(n)}$. As we will see in the next section, the equations of motion impose 
\begin{align}\label{solspace}
U^A&=U^A_0+r\,P^A+ \cO(r)^2\,, \qquad V=V_0+r\,V_1+ \cO(r)^2
\end{align}
where $V_0$ is chosen to be zero to describe a null hypersurface at $r=0$, and we further impose the boundary condition $U^A_0=0$. Indeed, as shown in \cite{Adami:2021nnf}, allowing for a non-vanishing $U^A_0$ does not yield a new independent charge and can then be safely set to zero. In case of a black hole horizon, $V_1$ is related to the surface gravity $\kappa$ by $V_1=2\kappa$.

We decompose $q_{AB}$ as an unimodular metric plus its determinant 
\begin{equation}\label{Omegasplit}
    q_{AB}(v,x^A)=\sqrt{q(v,x^A)}\, \bar q_{AB}(v,x^A)\text{ with }\det \bar q_{AB}=1 \,.
\end{equation} The expansion and the intrinsic shear of the null hypersurface are defined by 
\begin{equation}
\theta=\partial_v\ln\sqrt{q} \,,\quad \theta_{AB}=\partial_v\bar q_{AB} 
\end{equation} respectively. Notice that, in general, the intrinsic shear $\theta_{AB}$ does not vanish and contains genuine gravitational degrees of freedom, which is a key difference compared to the analysis at null infinity where this tensor is set to zero by Einstein's equations. It is useful to introduce the following derivative operator
\begin{equation}
    \hat D^{(m)}_A=D_A+ m P_A 
\end{equation} where $D_A$ is the Levi-Civita connection for $q_{AB}$. Finally, we use the notation $(AB)$ for symmetrization of two tensor indices $AB$ (i.e. $A_{(AB)}=\frac12(A_{AB}+A_{BA})$), and $\langle AB\rangle$ for the symmetric trace free part.

\subsection{Resolution of Einstein's equations}

We want to solve Einstein's equations, $E^{\mu\nu}=0$, where $E^{\mu\nu}=-R^{\mu\nu}+\frac12 g^{\mu\nu}({R^\rho}_\rho-2\Lambda)$ and $\Lambda$ is the cosmological constant, which we keep arbitrary for now. They can be decomposed into hypersurface equations, involving partial derivative equations in $r$,  $E^{uu}=0$, $E^{uA}=0$, $E^{AB}\gamma_{AB}=0$, and in evolution in $v$ equations, $(E^{ AB}-\frac12\gamma^{AB}E^{CD}\gamma_{CD})=0$, $E^{vr}=0$, $E^{rr}=0$, $E^{rA}=0$.

We first impose the hypersurface equations. 
$E^{uu}=0$ is given by
\begin{equation}
    E^{uu}=\partial_r^2\ln\sqrt{\gamma}-\frac14\partial_r\gamma^{AB}\partial_r\gamma_{AB} =0 \,.
\end{equation}
Given \eqref{MetricTaylorexpansion}, this equation will algebraically fix the traces of $\chi^{(n\geq2)}$, for instance
\begin{align}
    \chi^{(2)}&=\frac14\chi_{AB}\chi^{AB}+ \frac18\chi^2 \,.
    \end{align}
Then $E^{uA}=0$ is a second order differential equation in $r$ for $U^A$. The two free functions are $U_0^A$ and $P_A$. We have chosen the boundary condition $U_0^A=0$ and kept $P_A$ free. The higher orders in $r$ are algebraically determined. For instance, we have
    \begin{equation}
    U^{(2)}_A=\frac12 D_B\chi^{BA}-\frac12 \chi^{AB}P_B-\frac14 \partial^A\chi-\frac12\chi P^A \,.
    \end{equation}
Similarly imposing $E^{AB}\gamma_{AB}=0$ is a second order differential equation in $r$ for $V$. We have chosen the boundary condition $V_0=0$ and kept $V_1$ free.  The higher orders $V_{(n)}$ are algebraically determined. For instance, we have 
\begin{equation}\label{eomV2}
    V_2 %= \frac{1}{4} \theta\,\chi - \frac{1}{2} R - \frac{1}{4}\chi_{AB}\theta^{AB}+ \frac{3}{4}P_AP^A
    = -\frac12 \left(\partial_v +\frac12(\theta+V_1)\right)\chi -\Lambda -\frac14 \chi^{AB}\theta_{AB}-\frac12 \hat D_A^{(-1/2)}P^A \, . 
\end{equation}
 
The equation $(E^{ AB}-\frac12\gamma^{AB}E^{CD}\gamma_{CD})=0$ gives the time evolution of $\chi_{AB}$ and of the higher orders $\chi^{(n\geq2)}_{AB}$. We only write explicitly the equations for $\chi_{AB}$ and $\chi^{(2)}_{AB}$: {}
\begin{align}
\label{eomforchiAB}
&\left(\partial_v+\frac12V_1+\frac32\theta\right)\chi^{AB}+\chi^{(A}_C\theta^{B)C}+\frac14\chi \theta^{AB}+\hat D_{(1/2)}^{\langle A}P^{B\rangle}=0  
\end{align} and
\begin{align}
&\left(\partial_v+V_1+\frac32\theta\right)\chi_{(2)}^{AB}+\chi_{(2)}^{C(A}\theta^{B)}_C+\frac1{16}\left(\chi^2+3\chi_{CD} \chi^{CD}\right) \theta^{AB}+ \hat D_{(1)}^{\langle A}U_{(2)}^{B \rangle}\nonumber \\ 
&+\frac32 \chi^{C \langle A}D^{B \rangle}P_C+\frac34 P^{\langle A}D_C\chi^{B \rangle C}+\frac34  P_C D^{\langle A}\chi^{B \rangle C}+\frac58\chi \hat D^{\langle A}_{(1/2)}P^{B \rangle}
\\&+\left(\frac34\Lambda-\frac58 R-\frac38 \chi_{CD} \theta^{CD}+\frac7{16}\theta\chi-\frac18D_AP^A+\frac{11}{16}P_AP^A\right)\chi^{AB}=0 \,, \nonumber
\end{align}  where $R$ is the curvature for the metric $q_{AB}$. Imposing $E^{vr}=0$ yields only one equation (as the rest is trivially satisfied when the previous equations are imposed):
    \begin{align}
 & \left( \partial_v + \theta+\frac12 V_1\right)    \chi  -R+\hat D^{(1/2)}_A P^A+2\Lambda=0 \,. \label{chi evolution}
\end{align}
This is also the case for $E^{rr}=0$ and $E^{rA}=0$ that give the Raychaudhuri and Damour equations, respectively: 
\begin{align}
&\left(\partial_v-\frac12V_1+\frac12\theta\right)\theta+\frac14 \theta_{AB}\theta^{AB}=0, \label{Raychau} \\
&\left(\partial_v+2\theta\right)P^A-\partial^A(V_1+\theta)+\hat D^{(1)}_B \theta^{AB}=0\,,\quad\left(\partial_v+\theta\right)P_A-\partial_A(V_1+\theta)+D_B \theta^B_{A}=0 \, .\label{Dam}
\end{align}

To summarize, the solution space is fully characterized by providing the following data:
\begin{equation}
\begin{split}
\label{solspacemetric}
&\text{3 functions of $(v,x^A)$}\qquad V_1(v,x^A), \theta_{AB}(v,x^A) ,  \\
&\text{2 functions of $(r,x^A)$}\qquad \Sigma_{n=2}^{\infty} r^n\chi^{(n)}_{AB}(x^A) , \\
& \text{9 functions of $(x^A)$}\qquad \bar q^0_{AB}(x^A)\,,q_0(x^A), \theta_0(x^A)\,, P^0_A(x^A)\,, \chi_0(x^A) 
, \chi^0_{AB}(x^A) \,.
\end{split}
\end{equation}
On-shell, these functions allow to reconstruct the whole metric around the null hypersurface at $r=0$. It is therefore a well-posed characteristic initial value problem. In Appendix \ref{sec:Kerr}, we provide the explicit form of the Kerr solution in this parametrization.

\subsection{Leading and subleading phase space}
\label{sec:Leading and subleading phase space in metric formalism}

In this section, we consider the leading and subleading phase space around a bulk null hypersurface in the metric formalism. More precisely, we compute the presymplectic potential, the residual gauge diffeomorphisms, and the Barnich-Brandt charges \cite{Barnich:2001jy,Barnich:2003xg} (see also \cite{Compere:2007az,Compere:2018aar,Ruzziconi:2019pzd}). While the leading phase space has been investigated in great detail in the existing literature (see e.g. \cite{Adami:2021nnf}), the analysis of the subleading phase space is new and, as we shall see in Section \ref{sec:canonical symmetries at the horizon}, exhibits a structure similar to that of the leading radiative phase space at null infinity. Let us mention that the subleading phase space at null infinity was studied in \cite{Godazgar:2018vmm,Godazgar:2018dvh}, where it was shown to capture interesting features such as the Newman–Penrose conserved quantities. Our analysis of the subleading phase space at finite distance presented here follows a very similar spirit.

\subsubsection{Symplectic potential} \label{sec:sympletic pot}

We evaluate the Einstein-Hilbert presymplectic potential \cite{Iyer:1994ys,Wald:1999wa}
\begin{equation}
    \Theta_{EH}^\mu= \frac1{16\pi G} \sqrt{-g} \left( g^{\nu\rho}\delta \Gamma_{\nu\rho}^\mu -g^{\mu\nu}\delta \Gamma_{\nu\rho}^\rho\right)
    \label{EH pres pot}
\end{equation} for the metric discussed in the previous section. This object is obtained from the variation of the Einstein-Hilbert action, by keeping track of boundary terms. It is a one-form on the field space, as it involves a field variation $\delta$. The presymplectic current $\omega^\mu_{EH}$ can simply be obtained by taking one more variation. When evaluated on a $r=\text{constant}$ hypersurface, the relevant component of the presymplectic potential, $\Theta^r$, can be expanded as
\begin{equation}
\Theta^r=\Theta^r_{(0)}+\Theta^r_{(1)}\,r+\cO(r^2)
\label{expansion Theta}
\end{equation}
where the leading term reads as
\begin{equation}
\Theta^r_{(0)}=    -\frac1{16\pi G}\sqrt{q}\left[ \frac12 \theta^{AB}\delta \bar q_{AB}+\delta(V_1+\theta) \right]  +\frac1{16\pi G}\delta\left[\sqrt{q} \,\theta\right] . \label{leading presymplectic pot}
\end{equation} This result agrees with previous literature, see e.g. \cite{Hopfmuller:2016scf,Adami:2021nnf}. In the terminology of \cite{Freidel:2022vjq,Ciambelli:2023mir}, the first and the second term are respectively associated with the spin $2$ and spin $0$ Carrollian momenta. The spin $1$ term does not appear in this expression since we set to zero the associated source $U_0^A = 0$ below \eqref{solspace}. Furthermore, the subleading term is given by
\begin{align}\label{subleading sympl pot metric}
\Theta^r_{(1)}&=\frac{\sqrt{q}}{32\pi G}
\Big[- \theta^{AB}\delta \chi_{AB}-\left((\partial_v\chi_{CD})q^{C\langle A}q^{B\rangle D}+(V_1-\theta)\chi^{AB}+\frac12\chi \theta^{AB}+2D^{\langle A}P^{B \rangle}\right)\delta \bar q_{AB}  \nonumber\\
&+\left(\left(2\partial_v+V_1+\theta)\right)\chi+4V_2+2D_AP^A+2\chi^{AB}\theta_{AB}-q^{AB}\partial_v\chi_{AB}\right)\delta\ln \sq +2P_A\delta P^A\Big] \nonumber \\
&-\frac{1}{32\pi G} \Big[\partial_v\left[ \chi \delta\sq\right]-\partial_A\left[2\sqrt q \delta P^A\right]+\delta\left[\sqrt{q} \left(4V_2+\left(2\partial_v+2V_1+\theta\right)\chi+4D_AP^A\right) \right] \,.
\end{align}
At this stage, we stress that we have not used the constraint equations of motion, but only the radial equations of motion prescribing the radial behavior of $V$ and $U^A$ from the one of $\gamma_{AB}$ in Equation \eqref{MetricTaylorexpansion}. If we were to go fully on-shell, we would find that the subleading potential can be written only in terms of the Iyer-Wald ambiguities \cite{Iyer:1994ys,Wald:1999wa} as %\footnote{On-shell subleading potential \eqref{subleadingonshellmetric} is\begin{align1}    \Theta^r_{(1)} & = \frac1{8\pi G} \left(\partial_v\left[\sqrt{q} \left( \sigma_0 \bar m^A_0\delta \bar m_A^0-\rho_0 m^A_0\delta \bar m_A^0 \right) \right]  -\delta\left[\sqrt{q} \left(\tau_0\bar\tau_0-\rho_0 (\gamma_0+\bar \gamma_0)\right) \right]\right)+c.c.\\ &+\frac1{8\pi G} \sqrt{q} \,\eth\left[\bar\tau_0 (m^A_0\delta\bar m_A^0+\bar m^A_0\delta m_A^0)\right]+c.c. \end{align1}where the second line is a total derivative on the sphere. 
%Useful identity ? \begin{equation}-\left((\partial_v\chi_{CD})q^{C\langle A}q^{B\rangle D}+(V_1-\theta)\chi^{AB}\right)\delta \bar q_{AB}   = 4(\partial_v +\bar\lambda_0-\lambda_0-2(\gamma_0+\bar\gamma_0))\sigma_0\bar m_A^0\delta \bar m^A_0  +cc \end{equation}}
\begin{align}\label{subleadingonshellmetric}
\Theta^r_{(1)}&=\frac{1}{16\pi G}\left(\partial_v\left[\frac12\sqrt q q^{AB}\delta \chi_{AB}-\frac12\chi\delta \sqrt q\right]- \delta\left[ \frac{\sqrt{q}}2\left(P_AP^A +\chi V_1 \right)\right]-\partial_A\left[P^A\delta \sqrt q\right]\right) ,
\end{align}
in agreement with the general results of \cite{McNees:2023tus,McNees:2024iyu}. As we shall see later, it will be useful to consider the subleading presymplectic structure without imposing the constraint equations.

\subsubsection{Residual diffeomorphisms and charges}\label{app:charges}
\paragraph{Residual diffeomorphisms}
The residual gauge diffeomorphisms preserving the line element \eqref{metricGNC} and the boundary conditions $V = \mathcal{O}(r) = U^A$ are generated by
\begin{equation}\label{residualsymm}
\begin{split}
\xi^v&=f(v,x^A)\,, \quad\xi^A=Y^A(x^A) -r\,q^{AB}\partial_B f + \frac{r^2}2 \left(\chi^{AB}+\frac12q^{AB}\chi\right)\partial_A f+\cO(r^3)\,,\\
\xi^{r}&=-r\,\partial_v f+\cO(r^2)
\end{split}
\end{equation}
where the subleading orders in $r$ are fully determined by the leading functions $f$ and $Y^A$. 
These residual gauge transformations act on the leading order solution space as follows:
\begin{equation}
    \begin{split}
\delta_\xi q_{AB}&=(\pounds_Y + f \partial_v) q_{AB} \,,\qquad 
%%%%%%%%%%%%%%%%%%%%%%%%%%%%
\delta_\xi \ln\sqrt{q}=D_AY^A + f\partial_v\ln\sqrt{q} \,,\\
%%%%%%%%%%%%%%%%%%%%%%%%%%%%
\delta_\xi \theta_{AB}&=(\pounds_Y + f \partial_v+\partial_vf)\theta_{AB}\,,\qquad
%%%%%%%%%%%%%%%%%%%%%%%%%%%%
\delta_\xi \theta=\partial_v(f\,\theta)+Y^A\partial_A\theta \,, \\
%%%%%%%%%%%%%%%%%%%%%%%%%%%%
\delta_\xi V_1&=(\pounds_Y+f\partial_v+\partial_vf) V_1+2\partial_v^2 f \,, \\
\delta_\xi \chi_{AB}&=(\pounds_Y+f\partial_v-\partial_vf)\chi_{AB}  - 2\hat D^{(1)}_{\langle A}\partial_{B\rangle} f \,,\\
%%%%%%%%%%%%%%%%%%%%%%%%%%%%
\delta_\xi \chi&=(\pounds_Y+f\partial_v-\partial_vf)\chi- 2D^2 f -2P_A \partial^A f \,, \\
\delta_\xi P^A&=(\pounds_Y+f\partial_v)P^A+(V_1-\theta)\partial^Af- \theta^{AB}\partial_Bf-2\partial^A\partial_vf \,, \\
\delta_\xi \chi^{(2)}_{AB}&= (\pounds_Y+f\partial_v-2\partial_vf)\chi^{(2)}_{AB}  - \chi^C_{\langle A}\hat D^{(1)}_{C}\partial_{B\rangle} f -\frac12 \chi_{AB} P^C\partial_Cf \\
&
-2D_C\chi^{C}_{\langle A}\partial_{B\rangle}f
%-\chi_{C<A}P^C\partial_{B>}f
+\partial_{\langle A}\chi\partial_{B\rangle}f-\frac12\chi D_{\langle A}\partial_{B\rangle} f \,. 
    \end{split}
\end{equation}

\paragraph{Surface charges} The Barnich-Brandt co-dimension $2$ form $k_\xi^{\mu\nu}$ \cite{Barnich:2001jy,Barnich:2003xg} is a one-form on the field space which can be related to the above presymplectic current by the on-shell closure condition, $\partial_\nu k^{\nu\mu}_\xi = i_{\delta_\xi} \omega_{EH}^\mu$. Evaluating this co-dimension $2$ form on a constant $(v,r)$-surface, the relevant component is $k^{vr}_\xi$ and the (infinitesimal) surface charge reads as
\begin{equation}
    \ndelta H_\xi = \oint k^{vr}_\xi, \qquad \oint =\int d^2x \sq \, .
\end{equation} The notation $\ndelta H_\xi$ indicates that this object is still a one-form on the phase space which is generally not integrable (i.e. not $\delta$-exact \cite{Barnich:2007bf,Ruzziconi:2020wrb}). For a surface close to the horizon, we can expand the co-dimension $2$ form as
\begin{equation} \label{rad exp k}
k^{vr}=k^{vr}_0+r\, k_1^{vr}+\mathcal O(r^2) \, .
\end{equation} The leading order charge is given by 
\begin{equation}\label{leadingBB}
16\pi G   k_0^{vr}= Y^A\,\delta\left(\sqrt{\Omega} \,P_A\right)+ \left( f \,(V_1+\theta) +4 \partial_vf\right)\delta\sq -\frac{\sq}2 f\,\theta^{AB}\delta\bar q_{AB}+\partial_v\left(-2 f\,\delta \sq \right) \, .
\end{equation} The charge aspects appearing in this expression obey Raychaudhuri \eqref{Raychau} and Damour \eqref{Dam} evolution equations and have been studied in great details in previous literature \cite{Hawking:2016msc,Hawking:2016sgy,Carlip:2017xne,Haco:2018ske,Donnay:2015abr,Donnay:2016ejv,Chandrasekaran:2018aop,Grumiller:2019fmp,Adami:2020amw,Adami:2021nnf,Chandrasekaran:2021hxc,Odak:2022ndm,Odak:2023pga,Liu:2022uox}. In this work, we will instead focus on the subleading phase space corresponding to the order $r$ in the expansions \eqref{expansion Theta} and \eqref{rad exp k}. The associated charges are related to the ``radial evolution of the canonical charges'' discussed in \cite{Freidel:2024emv}.\footnote{The authors of \cite{Freidel:2024emv} discuss a ``spin-2'' charge that is related to the subleading charges presented here. We stress that it is different from the spin-2 charge $Q_2$ that we will encounter later in this work.} In that case, the co-dimension $2$ form can be computed as
\begin{align}
&16\pi G k_1^{vr} \nonumber\\
&=Y^A\delta Q_Y^A+f\delta Q_f-\frac12 f\sq \left(\delta\chi^{AB}\theta_{AB}+\chi\,\delta\theta \right) \nonumber \\
&+ \left[P^A\partial_Af+\frac12\chi(\partial_v+V_1)f+f\left( -\frac12\chi^{AB}\theta_{AB}+\frac12\partial_v\chi+2V_2+\hat D_A^{(-1)}P^A\right)\right]\delta\sq\\
%%%%%%%%%%%%%%%%%%%%
%& +\sq\Big[-\frac12 \chi^{AB}\delta \bar q_{AB}(\partial_v+V_1-\theta)f -\frac12 f\left((\partial_v\chi_{AB})^{\langle,\rangle} \delta \bar q_{AB}+\delta\chi^{AB}\theta_{AB}\right)\\
%%%%%%%%%%%%%%%%%%%%
%&-\frac14f\,\chi \left( \theta^{AB}\delta \bar q_{AB} +2\delta\theta \right) -\left(\partial^{\langle A}f\,P^{B \rangle}+f\,\hat D^{\langle A}_{(1/2)}P^{B \rangle} \right)\delta \bar q_{AB}\Big]
%%%%%%%%%%%%%%%%%%%%
&-\sq\Big[\frac12 \chi^{AB}(\partial_v+V_1-\theta)f+\partial^{\langle A}f\,P^{B \rangle} \nonumber\\
&\quad+\frac12f\left(  \partial_v\chi_{CD}\,q^{C\langle A}q^{B\rangle D}+\frac12\chi\, \theta^{AB}+2\hat D^{\langle A}_{(1/2)}P^{B \rangle}\right) \Big]\delta \bar q_{AB} \nonumber
%%%%%%%%%%%%%%%%%%%%
\end{align}
up to total derivative on the sphere and where %This expression can be simplify by using the equations of motion, we have \begin{align1}k_1^{vr}&=Y^A\delta   (Q_Y)_A+f\delta   Q_f\\\& +\frac1{16\pi G}\left(P^A\partial_Af+\frac12\chi(\partial_v+V_1)f+f\left( -\frac12R-\Lambda+\frac12\hat D_A^{1/2}P^A-\chi^{AB}\theta_{AB}\right)\right)\delta\sq\\%%%%%%%%%%%%%%%%%&+\frac{\sq}{16\pi G}\Big(-\frac12 \chi^{AB}\delta \bar q_{AB}\left(\partial_v+\frac12V_1-\frac12\theta\right)f -\frac12 f\delta\chi^{AB}\theta_{AB}\\%%%%%%%%%%%%%%%%%&-\frac14f\,\chi \left( \frac12\theta^{AB}\delta \bar q_{AB} +2\delta\theta \right) -\left(\partial^{\langle A}f\,P^{B \rangle}-\frac12 f\,\hat D^{\langle A}_{1/2}P^{B \rangle} \right)\delta \bar q_{AB}\Big)\end{align1}
%with \begin{align} ( Q_Y)_A&=\frac{\sq}{16\pi G}\left( D^B\chi_{AB}-\frac12\partial_A\chi\right)  =\frac{\sq}{16\pi G}  \Big[ 2 (\Psi^0_1)_A-\frac12\left(\chi_{AB}-\frac12 q_{AB}\chi\right)P^B\Big]\\Q_f&=\frac{\sq}{16\pi G}\Big(\frac12\theta\chi-R+2\Lambda \Big)=\frac{\sq}{16\pi G}\delta\Big[4\text{Re}\Psi_2^0-\frac12\chi^{AB}\theta_{AB}-\frac43\Lambda \Big] \end{align}where $(\Psi^0_1)_A$ is defined via $\Psi_1^0=(\Psi^0_1)_A m^A_0$. 
\begin{align}Q_Y^A&=\sq\left( \chi \,P^A+\chi^{AB}P_B+2U^A_2\right) \,,\quad 
Q_f=-\sq\left(\left(\partial_v+\frac12(\theta+V_1) \right)\chi+\hat D_A^{(1/2)}P^A  \right) \, .
\end{align} The charge aspects appearing in these subleading expressions are now related to the components of the Weyl tensor $\text{Re}\Psi_2^0$ and $\Psi_1^0$ which will be discussed in great details in the next section (see Equation \eqref{dictionary Weyl scalars}). On-shell we have 
\begin{align} ( Q_Y)_A&=\frac{\sq}{16\pi G}\left( D^B\chi_{AB}-\frac12\partial_A\chi\right)  =\frac{\sq}{16\pi G}  \Big[ 2 (\Psi^0_1)_A-\frac12\left(\chi_{AB}-\frac12 q_{AB}\chi\right)P^B\Big],\\Q_f&=\frac{\sq}{16\pi G}\Big(\frac12\theta\chi-R+2\Lambda \Big)=\frac{\sq}{16\pi G}\Big[4\text{Re}\Psi_2^0-\frac12\chi^{AB}\theta_{AB}-\frac43\Lambda \Big], \end{align}where $(\Psi^0_1)_A$ is defined via $\Psi_1^0=(\Psi^0_1)_A m^A_0$.   Intriguingly, readers familiar with the leading phase space analysis at null infinity may already notice the resemblance between these charge aspects and those found at null infinity. We will return to this observation in Section \ref{sec:From null infinity to the horizon}, where we discuss the map from null infinity to the horizon. Note that the subleading charge aspects described above will be related to the lower-spin charges ($s = 0,1$) of the infinite tower of spin-$s$ charges constructed in Section \ref{sec:Subleading tower of surface charges}, see also Appendix \ref{sec:Diffeomorphism interpretation}.

\section{From metric to Newman-Penrose formalism}
\label{sec:From metric to Newman-Penrose formalism}

In this section, we express the results obtained above in the Newman-Penrose (NP) formalism \cite{Newman:1961qr} (see also \cite{Penrose:1984uia,Chandrasekhar:1985kt,Newman:2009} for reviews), by carefully choosing a null tetrad. This allows us to build meaningful quantities out of the solution space, and identify a notion of transverse radiation through a null hypersurface. In particular, the present analysis extends the work of \cite{Liu:2022uox} in NP formalism by including intrinsic shear on the hypersurface, sourcing the radiation. We then show that the Bianchi identities at the horizon can be written in a Weyl-covariant form using appropriate Geroch-Held-Penrose (GHP) operators \cite{Geroch:1973am,Penrose:1985bww}.

\subsection{Construction of the Newman-Unti tetrad}
\label{sec:Construction of the Newman-Unti tetrad}

To use the Newman-Penrose formalism, one needs to choose a null tetrad. Following \cite{Geiller:2024bgf}, we discuss two choices: the Bondi tetrad and Newman-Unti tetrad (both compatible with the gauge-fixed metric \eqref{metricGNC}). The former has the advantage to be written in a closed form in terms of local functions the metric \eqref{metricGNC}. The latter instead has more vanishing spin coefficients \cite{Newman:1962cia}: 
\begin{equation}\label{NUtetraddef}
   \epsilon=\pi=\kappa =0\,,
\end{equation} which becomes very handy when discussing equations of motion. Spin coefficients can be computed from the tetrad $e_i$ and the metric $g_{\mu\nu} = e_\mu^i e_\nu^j \eta_{ij}$ through
\begin{equation}
    \gamma_{ijk}=\eta_{il} e_j^\mu e_k^\nu \nabla_\mu e^l_\nu
\end{equation}  where $\nabla_\mu$ is the Levi-Civita connection, $\epsilon=\frac12(\gamma_{121}-\gamma_{341})$, $\pi =- \gamma_{241}$ and $\kappa = \gamma_{131} $. As in \cite{Geiller:2024bgf}, we will write down the Bondi tetrad before performing an internal Lorentz transformation to reach the Newman-Unti tetrad. 

We follow the convention of \cite{Geiller:2022vto,Geiller:2024bgf}. In terms of the parametrization of the line element \eqref{metricGNC}, the Bondi tetrad $\tilde e_i=(\tilde \ell,\tilde n,\tilde m, \bar{\tilde m})$ reads as
\begin{equation}
\tilde \ell:=\partial_r\,, \quad \tilde n:=-\left(\partial_v+\frac V2 \partial_r+U^A\partial_A\right)\,,\quad 
\tilde m:=m^A\partial_A
\end{equation}
with 
\begin{align}
&m^A=\sqrt{\frac{\gamma_{\theta\theta}}{2\gamma}}\left(\frac{\sqrt{\gamma}+i\gamma_{\theta\phi}}{\gamma_{\theta\theta}}\,\delta^A_\theta-i\delta^A_\phi\right)
\end{align}
and we have $\tilde \ell^\mu \tilde n_\mu=-1=-\tilde {\bar{m}}_\mu \tilde m^\mu $, such that $g^{\mu\nu}=-\tilde \ell^\mu\tilde n^\nu-\tilde \ell^\nu\tilde n^\mu+\tilde {\bar{m}}^\mu \tilde m^\nu+\tilde {\bar{m}}^\nu \tilde m^\mu$.
 We compute the spin coefficients. We have a null vector $\ell$ which is geodesic, affinely parametrized and the gradient of a field, implying respectively 
\begin{equation}\label{tetraBondicond}
   \tilde \kappa=0\,,\quad  \tilde\epsilon+\bar {\tilde  \epsilon}=0\,,\quad \bar{ \tilde\rho}= \tilde\rho =\gamma_{134}\,,\quad \tilde \tau=\bar{\tilde\alpha}+\tilde \beta \, . 
\end{equation}
However $(\tilde n,\tilde m,\bar{\tilde m})$ is not parallelly transported along the geodesics since the following spin coefficients are non vanishing:
\begin{align}   
   \tilde \pi&=\left(\frac1{2}\gamma_{AB}\partial_rU^B \right)\tilde {\bar{m}}^A\,,\quad \tilde\epsilon=\frac18\partial_r\gamma_{AB}(\tilde{{m}}^A\tilde{{m}}^B-\tilde{\bar{m}}^A\tilde{\bar{m}}^B) .
\end{align}
This means that it does not satisfy the Newman-Unti (NU) defining criteria \eqref{NUtetraddef}. 

Following \cite{Geiller:2024bgf}, we use an internal Lorentz transformation of the null tetrad to reach the NU tetrad $e_i=(\ell,n,m,\bar m)$,
\begin{equation}\label{NUtetradLorentz}
    \ell:=e^{\theta_1}\tilde \ell\,, \quad n:=e^{-\theta_1}\tilde n+a\bar ae^{\theta_1}\tilde \ell+\bar a e^{i\theta_2}\tilde m+ae^{-\theta_2}\bar{\tilde m} \,, \quad m:= e^{i\theta_2}\tilde m +a e^{\theta_1}\tilde\ell
\end{equation}
with 
\begin{align}
    \theta_1& =0 \,,\quad 
   \partial_r \theta_2=-i(\tilde \epsilon-\bar{\tilde \epsilon}) \,,\quad 
   \partial_r \bar a =e^{-(\theta_1+i\theta_2)}\tilde \pi .
\end{align}
The fact that there are integrals over the radial coordinates $r$ (of non-total derivative terms) shows that we cannot write closed local expressions using the parametrization \eqref{metricGNC}. Using the expansion \eqref{MetricTaylorexpansion}, the NU tetrad is asymptotically given by
\begin{align}\label{NUtetrad}
    \ell^\mu&%=\tilde \ell^\mu 
    = \partial_r\,,\quad 
    n^\mu%=\tilde n^\mu+a\,\bar a \,\tilde \ell^\mu+e^{i\theta_2} \bar a\, \tilde m+e^{-i\theta_2} a \, \tilde  {\bar m}
    =-\partial_v-\frac r2\left(V_1\partial_r+P^A\partial_A \right) +\mathcal O(r)^2 \,,\quad
    m^\mu%=e^{i\theta_2}\tilde m+a \, \tilde \ell^\mu 
    = m_0^A\partial_A+ \mathcal O(r)
\end{align}
with \begin{equation}
m_0^A=\sqrt{\frac{q_{\theta\theta}}{2q}}\left(\frac{\sqrt{q}+iq_{\theta\phi}}{q_{\theta\theta}}\,\delta_\theta^A-i\delta^A_\phi\right) .
\label{transverse dyad}
\end{equation}
We can now verify that \eqref{NUtetraddef} is satisfied. In the following, we work exclusively with the Newman-Unti tetrad.  

\subsection{Dictionary from metric to NP}

In terms of the metric solution space discussed in Section \ref{sec:Phase space in metric formalism}, and using the Newman-Unti tetrad \eqref{NUtetrad}, the non-vanishing spin coefficients are
\begin{subequations} \label{dico spin coeff}
\begin{align}
\sigma&= \gamma_{133}=\frac12e^{2i\,\theta_2}(\partial_r\gamma_{AB} )m^Am^B=\frac12\chi_{AB}m_0^Am_0^B+\mathcal O(r) ,\label{shear expansion}\\
%%%%%%%%%
\alpha&=\frac{1}{2}(\gamma_{124}-\gamma_{344})=\frac{1}{2}\left( D_A+\frac12P_A\right){\bar{m}}_0^A+\mathcal O(r),\\
%%%%%%%%%
\beta&=\frac{1}{2}(\gamma_{123}-\gamma_{343})=-\frac{1}{2}\left( D_A-\frac12P_A\right){{m}}_0^A+\mathcal O(r),\\
%%%%%%%%%
\gamma
&=\frac{1}{2}(\gamma_{122}-\gamma_{342})=-\frac14\left( V_1+\frac12 \theta_{AB}({{m}}_0^A{{m}}_0^B-{\bar{m}}_0^A{\bar{m}}_0^B) \right)+\mathcal O(r), \label{gamma coefficient}\\
%%%%%%%%%
\nu
&=\gamma_{422}=\frac r2\left( \partial_vP_A-\partial_AV_1\right) \bar m^A_0+ \mathcal O(r)^2,\\
%%%%%%%%%
\mu
&=\gamma_{423}=\frac12\theta+\frac r4 \left(\left(\partial_v+\frac12V_1\right)\chi+\hat D_A^{(1/2)}P^A+i\,\epsilon^{AB}\,D_AP_B\right)+\mathcal O(r)^2,\\
%%%%%%%%%
\lambda&=\gamma_{424}=\frac12\theta_{AB}\bar m_0^A\bar m_0^B+\mathcal O(r),
\label{lambda expansion}\\
%%%%%%%%%
\tau& =\gamma_{132}%=\left(\partial_A\beta-\f{1}{2}e^{-2\beta}\gamma_{AB}\partial_rU^B\right)\hat{m}^A
=\frac12P_Am^A_0+\mathcal O(r),\\
%%%%%%%%%
\rho&=\gamma_{134}=\frac{1}{2}\partial_r\ln\sqrt{\gamma}
=\frac{1}{4}\chi+\mathcal O(r).
\end{align}
\end{subequations}
We introduce the notation 
\begin{equation}\label{conventionspincoef}
\gamma_{\{...\}}=(\gamma_{\{...\}})_0+(\gamma_{\{...\}})_1\, r  +\cO(r^2) \, .
\end{equation}
In particular, we have $\nu_0=0$, $\mu_0=\bar\mu_0$, 
and it is useful to write down the inverse expressions  
\begin{equation}
\begin{split}
\chi^{AB}&=2(\bar\sigma_0 m_0^Am_0^B+\sigma_0\bar m_0^A \bar m_0^B )\,,\quad \theta^{AB}=2(\lambda_0 m_0^Am_0^B+\bar\lambda_0\bar m_0^A \bar m_0^B)\, , \\P^A&=2 (\bar\tau_0\,m^A_0  +\tau_0\,\bar m^A_0 )\,, \quad q_{AB}=m^0_A\bar m^0_B+\bar m^0_A m^0_B \, . 
\end{split}
\end{equation}

The components of the Weyl tensor $W_{\mu\nu\rho\sigma}$ are denoted by
\begin{align}
\Psi_0&=-W_{\mu\nu\rho\sigma}\ell^\mu m^\nu \ell^\rho m^\sigma \,, \quad\Psi_1=-W_{\mu\nu\rho\sigma}\ell^\mu n^\nu \ell^\rho m^\sigma\,, \quad \Psi_2=-W_{\mu\nu\rho\sigma}\ell^\mu m^\nu \bar m^\rho n^\sigma ,\\
\Psi_3&=-W_{\mu\nu\rho\sigma}n^\mu \bar m^\nu n^\rho \ell^\sigma\,, \quad \Psi_4=-W_{\mu\nu\rho\sigma}n^\mu \bar m^\nu n^\rho \bar m^\sigma \, .
\end{align} and correspond to the Weyl scalars. Expanding in $r$ around the null hypersurface, we have
\begin{align}
    \Psi_n&=\Psi_n^0 +r\, \Psi_n^1+\mathcal O(r)^2\,,\quad n=0,...,4 \, .
\label{no Peeling}
\end{align} A crucial difference compared to null infinity is that the Weyl scalars do not peel at finite distance: each $\Psi_n$ starts at order $r^0$. In terms of the metric solution space, the coefficients in the expansion read as 
\begin{subequations} \label{dictionary Weyl scalars}
\begin{align}
\Psi^0_0& =\left(\chi^{(2)}_{AB}-\frac14\chi\,\chi_{AB}\right) m_0^Am_0^B \,,\quad   \Psi^1_0= \left( 3\chi^{(3)}_{AB}-\chi\,\chi_{AB}^{(2)}+\frac14\chi^2\chi_{AB}\right)m_0^Am_0^B , \label{Psi0n}   \\
\Psi^0_1&=\frac14\left(\left(\chi_{AB}-\frac12 q_{AB}\chi\right)P^A+2D_A\chi^{A}_B-\partial_B\chi\right)m^B_0 ,\\
%%%%%%%%%%%%%%%%%%%%%%%%%%%%%%%%%%%%%
\Psi^0_2
%&=\frac1{12}\left(R+\chi_{AB}\theta^{AB}+D_AP^A-2V_2+2P_AP^A+\partial_v\chi+\frac12 \chi V_1\right)+\frac{i}8\left( 2D_AP_B+\chi_{AC}\theta^{C}_B\right)\epsilon^{AB} \nonumber  \\
&=\frac1{4}\left(R+\frac12\chi_{AB}\theta^{AB}-\frac12\theta\chi-\frac23\Lambda\right)+\frac{i}8\left( 2D_AP_B+\chi_{AC}\theta^{C}_B\right)\epsilon^{AB} ,\\
%%%%%%%%%%%%%%%%%%%%%%%%%%%%%%%%%%%%%
\Psi^0_3
%&=\frac1{4}\left(-D_A\theta^A_B+\partial_B(\theta-V_1)+\theta_{AB}P^A+\partial_vP_B \right) \bar m_0^B  \nonumber \\
&=\frac1{2}\left(-\left(D_A-\frac12P_A\right)\theta^A_B+\left(\partial_B-\frac12 P_B\right)\theta \right) \bar m_0^B ,\\
%%%%%%%%%%%%%%%%%%%%%%%%%%%%%%%%%%%%%
\Psi^0_4&=\frac12\left(\partial_v\theta_{AB}-\frac12 V_1 \theta_{AB} \right)\bar m_0^A \bar m_0^B . 
\end{align}
\end{subequations} 
Each coefficient $\Psi^n_0$ in the radial expansion of $\Psi_0$ corresponds to a coefficient in the expansion of $\gamma_{AB}$ \eqref{MetricTaylorexpansion} starting at order $2$ (together with contributions of leading terms with respect to that order), schematically $\Psi_0^n\leftrightarrow \chi_{AB}^{(2+n)}m^A_0m^B_0+...$ (where the dots indicate possible leading contributions), see e.g. \eqref{Psi0n}. 

As we shall discuss later, each of the $\Psi^0_n$ has some nice Weyl-covariance properties and represents meaningful physical quantities near the horizon. In particular, by analogy with null infinity, it is tempting to identify $\Psi^0_4$ with the transverse radiation going through the horizon. Indeed, this object encompasses all the notions of radiation discussed in previous works, see e.g. \cite{Adami:2021nnf}. We will further justify this definition in Section \ref{sec:Subleading tower of surface charges} by showing that there exists a tower of conserved quantities for a self-dual subsector of gravity when $\Psi_4^0 =0$. However, despite these good properties, we will see in Section \ref{sec:Self-dual Kleinian Taub-NUT black holes} that stationarity does not necessarily imply $\Psi^0_4 = 0$. Hence, it seems that $\Psi^0_4$ contains additional information, and this notion of radiation would need to be refined to obtain an exact characterization of radiation at finite distance and nothing else.

As an illustration of the above discussion, in Appendix \ref{sec:Kerr}, we provide the explicit expression of the Weyl scalars at the horizon of a Kerr and Kerr-AdS black hole. In particular we have that $\Psi_4^0=0=\Psi_3^0$ and $\Psi_2^0\neq0$ and $\Psi_1^0\neq0$. The latter encode the mass and the rotation parameters, respectively.

\subsection{Solution space in NP}

In the previous section, we translated the solution space from metric to NP formalism by making a choice of null tetrad. In this section, we discuss the derivation of the solution space directly in NP formalism in the spirit of \cite{Newman:1962cia,Barnich:2016lyg,Barnich:2019vzx,Liu:2022uox}. More precisely, we want to understand the free data parametrizing the solution space that allow us to reconstruct the whole expansion and define the characteristic initial value problem. The NP equations are organized into metric equations, spin-coefficient equations and Bianchi identities. In Appendix \ref{sec:Relevant equations}, we write a few of these equations, and refer the reader to \cite{Penrose:1985bww,Chandrasekhar:1985kt,Newman:2009} for more details. 

We start with the NU tetrad $(L,N,M,\bar M)$ (at this stage, we consider it off-shell and do not connect it with the solution space in metric formalism):
\begin{align} \label{NU tetrad off shell}
    L^\mu&
    = \partial_r\,,\quad 
    N^\mu
    =-\partial_v-U(v,r,x^A)\partial_r-X^A(v,r,x^A)\partial_A\,,\quad
    M^\mu
    = M^A\partial_A+ \omega(v,r,x^A) \partial_r \, .
\end{align}
We impose \eqref{NUtetraddef} and \eqref{tetraBondicond} from the onset, and assume Taylor expandability in $r$.
By consistency with the NU tetrad written asymptotically in \eqref{NUtetrad}, we take the following boundary conditions:
\begin{equation}\label{NUBC}
M^A=\cO(1)\,,\quad \omega=\cO(r)\,,\quad   U=\cO(r)  \,, \quad X^A=\cO(r) .
\end{equation}
This is more general that the boundary conditions discussed in 
\cite{Liu:2022uox}. 
Before imposing any equations, we start with a space parametrized by the tetrad functions $U,X^A,M^A,\omega$, the 8 spin coefficients (recall \eqref{NUtetraddef} and \eqref{tetraBondicond}) and the 5 Weyl scalars. At this stage, all are complex functions of all coordinates. For the spin-coefficients, we use the notation \eqref{conventionspincoef}. 

Starting with the radial metric equations (Eq. (42)-(45) in \cite{Newman:2009}), the subleading orders of $U,X^A,\omega,M^A$ are determined in terms of the spin coefficients $\text{Re}(\gamma),\tau,\sigma,\rho$. Moreover because of the boundary conditions \eqref{NUBC}, they are fully specified except for the leading term in $M^A$, denoted by $m^A_0$. Then the evolution equation for $m^A_0$ is determined in terms of the spin coefficients $\mu_0,\gamma_0,\lambda_0$. Equation (48) at leading order imposes $\mu_0$ to be real and Equation (46) puts $\nu_0=0$. The last of the metric equations fixes the combination $\bar \alpha-\beta$, which leaves us with their sum unspecified, given by $\tau$. 

We now  impose the radial spin-coefficient equations (Eq. (50) in \cite{Newman:2009}) determining the {subleading terms in} the spin coefficients. We have that $\lambda,\rho,\alpha$ are determined in terms of $\sigma,\beta,\mu,\rho_0,\alpha_0,\lambda_0$. Imposing the remaining radial equations, we find that the {subleading terms in} $\sigma$, $\tau$, $\beta$, $\gamma$, $\mu$, $\nu$ are determined in terms of the Weyl scalars $\Psi_0,\Psi_1,\Psi_2,\Psi_3$ and the leading spin-coefficients. Similarly the radial Bianchi identities (Eq. (52) in \cite{Newman:2009}) give the subleading components of $\Psi_4,\Psi_3,\Psi_2,\Psi_1$ in terms of $\Psi_0, \Psi_1^0, \Psi_2^0, \Psi_3^0, \Psi_4^0$ and the spin-coefficients $\lambda,\rho,\alpha$. At this stage we have a solution spanned by 
\begin{equation}
\begin{split}
& \Psi_0(v,r,x^A) \,,\\
& \text{Re}(\mu_0)(v,x^A)\,,{\text{Re}}(\rho_0)(v,x^A),\sigma_0(v,x^A),\tau_0(v,x^A),\lambda_0(v,x^A)\,,\gamma_0(v,x^A)\,,\\
&\Psi_1^0(v,x^A), \Psi_2^0(v,x^A), \Psi_3^0(v,x^A), \Psi_4^0(v,x^A) \,,\\
& m^A_0(x^A) \,.
\end{split} \label{free data}
\end{equation}

We now discuss the remaining equations constraining the leading coefficients. We impose (Eq. (53) in \cite{Newman:2009}) which gives the evolutions of $\Psi_0, \Psi_1^0, \Psi_2^0, \Psi_3^0$ in terms of $\Psi_4^0,\gamma_0,\tau_0,\sigma_0$. 
We impose (Eq. (51) in \cite{Newman:2009}) giving the evolution of $\lambda_0$ in terms of $\mu_0,\gamma_0,\tau_0,\Psi_4^0$. Moreover the last five equations in (51) constrain $\mu_0,\tau_0,\sigma_0,\rho_0$. Finally, we should satisfy the second to fourth equations in (51) which relate $\Psi^0_1,\Psi^0_2,\Psi^0_3$ to the spin coefficients. We are left with 
\begin{equation}
\begin{split}
& \Psi_0(r,x^A) , \\
&\text{Re}(\mu_0)(x^A), {\text{Re}}(\rho_0)(x^A),\tau_0(x^A),\sigma_0(x^A),\lambda_0(x^A) , \\
&\Psi_4^0(v,x^A)\,,\gamma_0(v,x^A) , \\
& m^A_0(x^A) . 
\end{split}\label{free data2}
\end{equation}

The solution space in NP formalism is consistent with the one in metric formalism, though encoded differently. In metric formalism, the characteristic initial value problem was summarized in
\eqref{solspacemetric}, and the relation to NP variables is summarized in Table \ref{tab:my_label} (there are two equivalent views of $\lambda_0$ and $\Psi_4$: either you consider $\lambda_0$ as an arbitrary function of $(v,x^A)$ that determines $\Psi_4^0$, or you view $\Psi_4^0(v,x^A)$ and $\lambda_0(x^A)$ as independent data determining $\lambda_0(v,x^A)$). 
\begin{table}[h]
    \centering
    \begin{tabular}{cc|c|c}
Number of &(real) functions of     &  Metric & NP \\
\hline
3&$(v,x^A)$     & $V_1$ & $\text{Re}(\gamma_0)$ \\
& & $\theta_{AB}$ & $ \lambda_0$ \\
\hline
2& $(r,x^A)$    & $\Sigma_{r=2}^{\infty} r^n\chi^{(n)}_{AB}$ & $\Psi_0$\\
\hline
9& $(x^A)$    &$\bar q^0_{AB},q_0$ & $m^A_0$\\
& &$ \theta_0$ & ${\text{Re}}(\mu_0)$\\
& &$ P^0_A$ & $\tau_0$\\
& &$\chi_0 $ & ${\text{Re}}(\rho_0)$\\
& &$\chi^0_{AB} $ & $\sigma_0$\\
    \end{tabular}
    \caption{Correspondence between the free functions in NP and in metric.}
    \label{tab:my_label}
\end{table}

Notice that from the above discussion in NP formalism, $\text{Im} (\gamma_0)$ is also a free datum in \eqref{free data2}. As noticed in \cite{Barnich:2019vzx}, at null infinity, this extra degree of freedom corresponds to the choice of leading order transverse dyad $(m_0, \bar m_0)$. To go from metric to first order formalism, we picked a particular choice of transverse dyad \eqref{transverse dyad}. Rotating the dyad as
\begin{equation}\label{rescaldyad}
    m_0 \to  e^{i b} m_0 , \qquad \bar{m}_0 \to e^{-i b}  \bar{m}_0 
\end{equation} ($b(v,x^A)\in \mathbb{R}$) changes the expression of $\text{Im} (\gamma_0)$ in \eqref{gamma coefficient}, more precisely we have $\text{Im} (\gamma_0)\to\text{Im} (\gamma_0)+\frac12\partial_vb$. Hence this extra degree of freedom in NP formalism will be invisible in {the} metric and only appears in the translation between the two formalisms. We can use this freedom later to set to zero the imaginary part of $\gamma_0$. The expression of the other spin coefficients in \eqref{dico spin coeff} and Weyl scalars \eqref{dictionary Weyl scalars} will remain the same provided we use the rescaled dyad \eqref{rescaldyad}.

Using the dictionary between metric and NP displayed in Equation \eqref{dico spin coeff}, we see that Equation \eqref{NPRaych} is the Raychaudhuri equation \eqref{Raychau}, Equations \eqref{NPDamour1} and \eqref{NPDamour2} reproduce the Damour equation \eqref{Dam}, and Equation \eqref{NPeqchi} is the evolution equation of $\chi_{AB}$ \eqref{eomforchiAB}. Hence, we conclude that the leading order analysis in the metric formalism corresponds to the spin-coefficient equations and not to the Bianchi identities, by contrast with the situation at null infinity.

\subsection{Covariantization}
\label{sec:Covariantization}

One of the advantages of the NP formalism is to trade tensors transforming under {diffeomorphisms} for simpler weighted scalars under null tetrad transformations. As we will explain in this section, this allows us to write complicated expressions in the solution space and evolution equations in a very compact and covariant way. We briefly review the definition of weighted scalars and corresponding GHP derivative operators \cite{Geroch:1973am}, based on Sections 4 and 5 of \cite{Penrose:1984uia}. We then show that the evolution equations \eqref{Bianchiradial} can be rewritten in a fully Weyl-covariant way.  

Consider the following transformations of the null tetrad preserving the normaliation conditions and the null directions of $\ell^\mu$ and $n^\mu$: 
\begin{equation}
    \ell^\mu \to \mathcal{V}  \ell^\mu , \quad n^\mu \to \mathcal{V}^{-1} n^\mu , \quad m^\mu \to e^{i b} m^\mu, \quad \bar{m}^\mu \to e^{-i b} \bar{m}^\mu 
    \label{transfo tetrad}
\end{equation} where $\mathcal{V}$ and $b$ are real. A scalar $\eta$ of spin weight $s$ and boost weight $w$ transforms as follows under \eqref{transfo tetrad}:
\begin{equation}
    \eta \to \mathcal{V}^w e^{i s b} \eta \, . 
\end{equation} It is also useful to introduce the weights $p = w+ s$ and $q = w-s$. Note that under complex conjugation, $\bar{\eta}$ has weights $\{ \bar{w}, \bar{s} \} =\{ w, -s \}$ and $\{ \bar{p}, \bar{q} \} =\{ q, p \}$. Examples of weighted scalars are the Weyl scalars $\Psi_n$, with weights $\{p=2(2-n),q=0\}$ for $n=4,3,2,1,0$, and the spin coefficients $\lambda: \{-3,1\}$ and $\sigma:\{ 3,-1 \}$. However, not all of the spin coefficients admit well-defined weights --- some exhibit anomalous terms in their transformation under \eqref{transfo tetrad}. 

We then define the differential operators associated with the null tetrad:
\begin{equation}
    D=\ell^\mu\partial_\mu \,, \quad\Delta=n^\mu\partial_\mu\,, \quad\partial=m^A\partial_A\,, \quad \bar\partial=\bar m^A\partial_A \, .
\end{equation}
In general, when acting on a weighted scalar $\eta$ of weights $\{p,q\}$, they do not produce a weighted scalar. To remedy this problem, one can correct the derivative operators using spin coefficients transforming anomalously under \eqref{transfo tetrad}. In the NU tetrad satisfying \eqref{NUtetraddef}, we define the GHP derivative operators
\begin{equation}
\begin{split}\label{weightopderivation}
\pthorn &=D\,,\qquad \pthorn '=\Delta +p  \gamma +q \bar \gamma \,, \\
\eth&= \partial +( p\, \beta+q\, \bar\alpha)  \,,\qquad
\bar \eth= \bar \partial +( p\, \alpha+q\,\bar \beta)  \, .
\end{split}
\end{equation} When acting on a scalar $\eta$ of weights $\{ p,q \}$, they produce weighted scalars: $\pthorn \eta: \{ p+1,q+1 \}$, $\pthorn' \eta: \{ p-1,q-1 \}$, $\eth \eta: \{ p+1,q-1 \}$ and $\bar{\eth} \eta: \{ p-1,q+1 \}$. For instance, in terms of spacetime derivative, we have
$D_{\langle A}P_{B\rangle}=2(\eth \tau_0\,\bar m_A^0 \bar m_B^0+\bar\eth \bar\tau_0 \, m_A^0 m_B^0)$ with $\tau_0:\{1,-1 \}$.

In addition to the transformations \eqref{transfo tetrad}, one can consider the rescaling of the tetrad
\begin{equation}
\label{rescaling tetrad}
    \ell^\mu \to \Omega^{-2} \ell^\mu, \quad n^\mu \to n^{\mu}, \quad m^\mu \to \Omega^{-1} m^\mu, \quad \bar{m}^\mu \to  \Omega^{-1} \bar{m}^\mu
\end{equation} (case (iv) in Eq. (5.6.26) of \cite{Penrose:1984uia}), associated with the Weyl rescaling of the metric $g_{\mu\nu} \to \Omega^2 g_{\mu\nu}$. Similarly to the above discussion, a Weyl scalar $\eta$ with Weyl weight $W$ is defined as a scalar transforming as 
\begin{equation}
    \eta \to \Omega^W \eta
\end{equation} under \eqref{rescaling tetrad}. This weight is invariant under complex conjugation. The weights are collectively denoted by $\{p, q, W \}$. The Weyl scalars admit well-defined Weyl weights:
\begin{equation}
        \Psi_n:\{2(2-n),0,n-5\} \, .
    \label{weyl weights Psi}
\end{equation}
Most of the spin coefficients are not Weyl weighted scalars at the exception of 
\begin{align}
&\lambda: \{-3,1,0\} \, ,   \qquad \sigma:\{ 3,-1 , -2\} \, .
\label{Weyl weights shear}
\end{align} 
By construction the tetrad elements are weighted objects. For instance, we have
\begin{align}
&m^A: \{1,-1,-1\} \, ,    \qquad \bar m^A:\{ -1,1 ,-1\}
\end{align} 
and $\sqrt{q}:\{0,0,2\}$. However, the derivative operators \eqref{weightopderivation} do not have well-defined Weyl weights. One can play the same game as above and use the spin coefficients that transform anomalously under \eqref{rescaling tetrad} to define new derivative operators in the NU tetrad:
\begin{equation}
\begin{split}
\pthorn_{\mathscr C}&=\pthorn+W\rho\,,\qquad  \pthorn'_{\mathscr C} =\pthorn'+ (W +p+q) \mu \,, \\
%\eth_{\mathscr C} &= \partial +p\, \beta_0+q\, \bar\alpha_0+(w- q) \, \tau_0 = \partial +(p+w- q)\, \beta_0+w\, \bar\alpha_0\\
\eth_{\mathscr C} &=\eth-(W+q) \, \tau \,,\qquad
\eth'_{\mathscr C} = \bar\eth \,.
\end{split}
\end{equation}
This is a subcase of Equation (5.6.36) of \cite{Penrose:1984uia} for the weight $w_0=-1$ and $ w_1=0$, and with the NU conditions \eqref{NUtetraddef}. The differential operators have weights $\{p,q,W\}$ given by
\begin{equation}
\pthorn_{\mathscr C}:\{1,1,-2\} \,, \quad 
\pthorn'_{\mathscr C}:\{-1,-1,0\} \,, \quad  \eth_{\mathscr C}:\{1,-1,-1\} \,, \quad \eth'_{\mathscr C}:\{ -1,1 , -1 \} \,. 
\end{equation}

Using the radial expansion of the tetrad \eqref{NUtetrad} and the spin coefficients with the convention \eqref{conventionspincoef}, the leading differential operators read as
\begin{subequations}
\label{covariant derivative operators}
\begin{align}
(\pthorn_{\mathscr C})^0&=\partial_r+W\rho_0 \, , \\
 (\pthorn'_{\mathscr C})^0 &=-\partial_v+p\,\gamma_0+ q \bar\gamma_0 +(W +p+q) \mu_0  \, , \label{thorn prime}\\
%\eth_{\mathscr C} &= \partial +p\, \beta_0+q\, \bar\alpha_0+(w- q) \, \tau_0 = \partial +(p+w- q)\, \beta_0+w\, \bar\alpha_0\\
(\eth_{\mathscr C})^0 &= \partial_0 +p\, \beta_0+q\, \bar\alpha_0-(W+q) \, \tau_0 = \partial_0 +(p-W- q)\, \beta_0-W\, \bar\alpha_0  \, ,\\
(\eth'_{\mathscr C})^0 &=\bar \partial_0 +( p\, \alpha_0+q\,\bar \beta_0)  \, ,
\end{align}
\end{subequations}
where $\partial_0=m_0^A\partial_A$.
In the following, we will always consider these leading derivative operators and we will drop the superscript `$0$' for brevity:
\begin{equation}
  \text{Notations:}\quad  (\pthorn'_{\mathscr C})^0 \to (\pthorn'_{\mathscr C}) , \quad (\pthorn_{\mathscr C})^0 \to (\pthorn_{\mathscr C}) , \quad (\eth_{\mathscr C})^0 \to (\eth_{\mathscr C}), \quad (\eth'_{\mathscr C})^0 \to (\eth'_{\mathscr C})  \, . 
\end{equation}
We also introduce the compact notation for the Weyl scalars
\begin{equation}
 Q_s=\Psi_{2-s}^0 \quad\text{ for }s=-2,-1,0,1,2 \, . 
 \label{Weyl scalars}
\end{equation} where the subscript $s$ denotes the spin. They have weights\footnote{More explicitly, we have:
\[
\begin{split}
&Q_{-2}=\Psi_4^0: \{-4 , 0 ,-1\}\,,\qquad
Q_{-1}=\Psi_3^0:\{-2, 0 , -2\}\,,\qquad
Q_{0}=\Psi_2^0:\{0 ,0 , -3\},\\
&Q_{1}=\Psi_1^0: \{2 , 0 ,-4\}\,,\qquad
Q_{2}=\Psi_0^0:\{ 4 , 0 , -5\} \, .
\end{split}
\] }
\begin{equation}
Q_s:\{2s,0,-(3+s)\}   \, .  
\end{equation}
With the above definitions, we can now rewrite the Bianchi evolution equations on a null hypersurface \eqref{eq:Bianchi evolution} compactly:
\begin{equation}
\boxed{
     \pthorn'_{\mathscr C} Q_s= \eth_{\mathscr C} Q_{s-1}-(s+1)\sigma_0\, Q_{s-2} \, .
}
\label{Bianchi evolution equations}
\end{equation} These equations are fully covariant under tetrad transformations \eqref{transfo tetrad} and Weyl rescalings \eqref{rescaling tetrad}. This last fact will be exploited in the next section to establish the relation with null infinity.  

Another example of equation of motion that can be rewritten in a Weyl covariant way is given \eqref{evolmbA}: $\pthorn'_{\mathscr C}\bar m_A^0=-\lambda_0  m_A^0$. Of course, not all the equations of motion are Weyl covariant. For instance, \eqref{NPeqchi} is
\begin{equation}
(\pthorn'_{\mathscr C}-\mu_0)\sigma_0-\eth \tau_0-\tau_0^2-\rho_0 \bar\lambda_0 =0 \, . 
\end{equation}
Here $\tau_0$ has well-defined weight $(p,q)=(1,-1)$ but no good Weyl rescaling property.

For later use, we compute commutators of some of the Weyl-covariant derivative operators \eqref{covariant derivative operators} acting on a weighted scalar $\eta$ of weights $(p,q,W)$. Using the metric equation and the spin coefficient evolution equations displayed in Appendix \ref{sec:Relevant equations}, we obtain
\begin{align1}
[\pthorn'_{\mathscr C},\eth_{\mathscr C}]\eta&= \bar\lambda_0  \eth'_{\mathscr C} \eta
-\eta\Big[(p+q) \partial_0\mu_0-(p-q-2W)\mu_0\tau_0\\&+(W\bar\partial_0+(q+3W)\alpha_0+(q-W)\bar\beta_0)\bar\lambda_0+(q+W)\partial_0(\gamma_0+\bar\gamma_0)
\Big] \, . 
\end{align1}
In particular, for the Weyl scalars \eqref{Weyl scalars}, we have 
\begin{align1} \label{commutation thornprimeEth}
&[\pthorn'_{\mathscr C},\eth_{\mathscr C}]Q_s \\
&= \bar\lambda_0  \eth'_{\mathscr C} Q_s
-Q_s\Big[2s \partial_0\mu_0-(6+4s)\mu_0\tau_0-(3+s)((\bar\partial_0+3\alpha_0-\bar\beta_0)\bar\lambda_0+\partial_0(\gamma_0+\bar\gamma_0))
\Big]   \\
&=\bar\lambda_0  \eth'_{\mathscr C} Q_s +(3-s)Q_s\eth'_{\mathscr C}\bar\lambda_0
-Q_s\Big[2s \bar Q_{-1}-(3+s)(2\mu_0\tau_0+2\bar\lambda_0\bar\tau_0+\partial_0(\gamma_0+\bar\gamma_0))
\Big]   
\end{align1}
where, in the second equality, we used 
\begin{align}
    \bar Q_{-1}&=\partial \mu_0-\mu_0\tau_0-\bar\partial_0 \bar\lambda_0-\bar\lambda_0(\alpha_0-3\bar\beta_0)=\partial \mu_0-\mu_0\tau_0-\eth'_{\mathscr C} \bar\lambda_0 \, . 
    %\\ Q_{-1}&=\bar \partial \bar \mu_0-\bar \mu_0\bar \tau_0-\partial_0\lambda_0-\lambda_0(\bar\alpha_0-3\beta_0)=\bar\eth \bar\mu_0-\eth \lambda_0=\bar\eth \bar\mu_0-\eth_{\mathscr C} \lambda_0-2\tau_0\lambda_0
\end{align}
Furthermore, we have
\begin{align1}\label{commbareththornprim}
[\pthorn'_{\mathscr C},\eth'_{\mathscr C}]\eta&= \bar\tau_0\pthorn'_{\mathscr C} \eta+\lambda_0  \eth_{\mathscr C} \eta
-\eta\Big[((2p+q+W) \bar\partial_0-(p+q)\bar\tau_0)\mu_0\\&+(-p\partial_0-(2p+q+W)\bar\alpha_0+(2p-q-W)\beta_0)\lambda_0
\Big] \, . 
\end{align1}

\section{From null infinity to the horizon}
\label{sec:From null infinity to the horizon}

\subsection{Weyl rescaling, Peeling theorem and recursion relations}
\label{sec:Weyl rescaling on phase space and recursion relations}

A drastic difference between the asymptotic spacetime boundary and a finite-distance null hypersurface is the role played by the cosmological constant: at infinity, it affects the leading part of the metric which dictates the nature of the spacetime boundary (either timelike ($\Lambda<0$), spacelike ($\Lambda>0$) or null ($\Lambda=0$)), while at finite distance, it appears only at subleading order. For the rest of the paper we will restrict ourselves to $\Lambda=0$ to discuss the interplay between null infinity ($\mathscr{I}$) and a finite-distance null hypersurface.

Null infinity can be treated as a null hypersurface at finite distance through Penrose's conformal compactification \cite{Penrose:1962ij, Penrose:1964ge,Newman:1966ub, Penrose:1986uia} by performing a Weyl rescaling on the bulk metric. Of course, it is not a generic null hypersurface: when imposing the leading order equations of motion, it corresponds to a weakly isolated horizon through the conformal compactification, see \cite{Ashtekar:2024mme, Ashtekar:2024bpi, Riello:2024uvs, Ashtekar:2024stm}. Here, following \cite{Ruzziconi:2025fct}, we will perform the conformal compactification off-shell and show how radial expansions at null infinity are mapped onto radial expansions at finite distance discussed in the previous sections.

We start from the radial expansion of the NU tetrad \cite{Newman:1962cia} near $\mathscr I \equiv \{ r_{\mathscr {I}} \to \infty \}$:
\begin{equation}
\begin{split}
    &\ell = -\partial_{r_\mathscr{I}}, \qquad n =(-1+ \mathcal{O}(r^{-1}_{\mathscr I})) \partial_v+\mathcal{O}(r_{\mathscr I}) \partial_{r_\mathscr{I}} + \mathcal{O}(r^{-1}_{\mathscr I})^A \partial_A, \\
    &m = \frac{1}{r_{\mathscr{I}}} m_0^A \partial_A + \mathcal{O}(r_{\mathscr{I}}^{-2}), \quad \bar{m} = \frac{1}{r_{\mathscr{I}}}\bar{m}_0^A \partial_A + \mathcal{O}(r_{\mathscr I}^{-2}) 
   \label{tetrad exapansion}
\end{split}
\end{equation} where $v$ is the null time along $\mathscr{I}$ and $q_{AB} = m_A^0 \bar{m}_B^0 + \bar m_A^0 m_B^0$ is the two-dimensional boundary metric, which is usually fixed on the phase space. Performing the conformal rescaling and using \eqref{rescaling tetrad}, we find 
\begin{equation}  \label{NU tetrad horizon}
\begin{split}
    &\ell \to \Omega^{-2}\partial_{r_\mathscr{I}} = \partial_r , \quad n \to \Omega^0 n=-\partial_v  + \mathcal{O}(r) ,\\
    &m \to \Omega^{-1}\Big[\frac{1}{r_{\mathscr{I}}} m^A_0 \partial_A + \mathcal{O}(r_{\mathscr I}^{-1}) \Big] =  m^A_0 \partial_A + \mathcal{O}(r), \\     &\bar{m} \to \bar{m}' = \Omega^{-1}\Big[\frac{1}{r_{\mathscr{I}}}\bar{m}_0^A \partial_A + \mathcal{O}(r_{\mathscr I}^{-1}) \Big] = \bar{m}_0^A \partial_A + \mathcal{O}(r) .
    \end{split}
\end{equation} 
We identified the conformal factor $\Omega \sim r^{-1}_{\mathscr{I}} \sim r$ with the radial coordinate $r \ge 0$ ($r=0$ is the locus of a finite-distance null hypersurface $\mathcal{H}$), and $v\in \mathbb{R}$ is now the null time along $\mathcal{H}$. By contrast with null infinity, the leading terms $m^A_0$ and $\bar{m}^A_0$ characterizing the intrinsic geometry of $\mathcal{H}$ are now part of the phase space and encode genuine radiative degrees of freedom. The near-horizon expansion \eqref{NU tetrad horizon} reproduces correctly \eqref{NU tetrad off shell} with \eqref{NUBC}.

Similarly, the $1/r_{\mathscr{I}}$ expansion of the Weyl tensor in asymptotically flat spacetime at null infinity ($\mathscr{I} \equiv r_{\mathscr{I}} \to \infty)$ is controlled by the Peeling theorem: 
\begin{equation}
    \Psi_n = \frac{\Psi_n^0}{r_{\mathscr I}^{5-n}} + \mathcal{O}(r_{\mathscr I}^{n-6}) \qquad n = 0,1, 2, 3, 4.
    \label{Peeling}
\end{equation} This is drastically different from the behavior at finite distance \eqref{no Peeling}. However, if one performs a Weyl rescaling with conformal factor $\Omega \sim r^{-1}_{\mathscr{I}} \sim r$, we have 
\begin{equation}
    \Psi_n \to \Psi'_n = \Omega^{n-5} \left[ \frac{\Psi_n^0}{r_{\mathscr I}^{5-n}} + \mathcal{O}( r_{\mathscr I}^{n-6})) \right] = \Psi^0_n + \mathcal{O}(r)
\end{equation} where we used the Weyl weights of $\Psi_n$ given in \eqref{weyl weights Psi}. Hence, the Peeling \eqref{Peeling} at null infinity consistently reduces to the Taylor expansion \eqref{no Peeling} at finite distance after Weyl rescaling. 

Furthermore, at null infinity, we have the following $1/r_{\mathscr{I}}$ expansion for the shear:
\begin{equation}
    \sigma = \frac{\sigma_0}{r_{\mathscr I}^2} + \mathcal{O}(r_{\mathscr I}^{-3}), %\qquad \lambda = \frac{\lambda^{\mathscr I}_1}{r} + \mathcal{O}(r)
\end{equation} where $\sigma_0$ is the asymptotic shear at $\mathscr{I}$. Therefore under Weyl rescaling, we have
\begin{equation}
    \sigma \to    \sigma' =\Omega^{-2}\frac{\sigma_0}{r_{\mathscr I}^2} + \mathcal{O}(r_{\mathscr I}^{-3}) = \sigma_0 + \mathcal{O}(r)
\end{equation} after using the Weyl weight of $\sigma_0$ in \eqref{Weyl weights shear}. Again, this expansion matches with the Taylor expansion \eqref{shear expansion} at finite distance. 

Finally, we consider the $1/r_\mathscr{I}$ expansion at null infinity 
\begin{equation}
    \lambda = \lambda_0 + \frac{\lambda_1}{r_{\mathscr{I}}} +\mathcal{O}(r^{-2}_{\mathscr I})
\label{expansion lamba}
\end{equation} The first term $\lambda_0$ vanishes because of Einstein's equations (the intrinsic shear of $\mathscr{I}$ vanishes). However, as emphasized earlier, we do not impose this on-shell condition prior to the conformal compactification. Since the Weyl weight of $\lambda$ is $0$, the rescaling simply gives 
\begin{equation}
    \lambda \to \lambda' = \lambda_0 + \lambda_1 r +\mathcal{O}(r^{2})
\end{equation} and we recover \eqref{lambda expansion}. However, the crucial difference is that $\lambda_0$ is no longer zero for a generic null hypersurface at finite distance.

In summary, we have a perfect matching between the radial expansions at null infinity and those found in the previous sections at the horizon through the off-shell conformal rescaling \cite{Ruzziconi:2025fct}. The second step in the identification between physics at null infinity and at the horizon will be achieved in Section \ref{sec:canonical symmetries at the horizon} by identifying the symplectic structures.

Based on the above considerations, it is legitimate to see which symmetry structure survives when going from null infinity to the horizon. We list below two key ingredients to identify the $Lw_{1+\infty}$ symmetries in the phase space at $\mathscr{I}$ from a spacetime perspective \cite{Freidel:2021ytz, Geiller:2024bgf,Kmec:2024nmu,Kmec:2025ftx}: 
\begin{itemize}

\item The recursion relations encoding the hierarchies of the integrable self-dual sector of gravity are the starting point of the analysis:
    \begin{equation} \label{full recursion at null infinity}
    \begin{split}
         -\partial_v Q_s & = \eth Q_{s-1}-(s+1)\sigma_0Q_{s-2} \\s&= -1, 0, 1, 2, 3, \ldots
         \end{split}
    \end{equation} 
     Each $Q_s$ for $s>-2$ corresponds to a charge aspect of spin weight $s$ and $Q_{-2} \equiv \Psi^0_4 = \partial_v \lambda_1$ is the radiation responsible for the non-conservation at $\mathscr{I}$. These recursion relations correspond to an infinite tower of flux-balance laws at null infinity. For $s=-2, \ldots, 2$, we have $Q_{s} = \Psi^0_{2-s}$ and Equation \eqref{full recursion at null infinity} follows from the Bianchi identities. For $s>2$, the $Q_s$'s appear in the $1/r_{\mathscr{I}}$ expansion of $\Psi_0$. Self-dual gravity and full gravity exhibit the same pattern only at leading orders near null infinity. In \cite{Geiller:2024bgf}, deviations with respect to the self-dual sector were observed for $s \ge 4$, and obtaining \eqref{full recursion at null infinity} from the full gravity theory requires a truncation of the Bianchi identities. The importance of the self-duality assumption to find the whole tower of charge aspects is particularly transparent in the top-down derivation using the twistor approach \cite{Kmec:2024nmu, Kmec:2025ftx}.

\item The Ashtekar-Streubel symplectic structure on the radiative phase space \cite{Ashtekar:1981bq} can be obtained from standard covariant phase space methods \cite{Iyer:1994ys} by pushing a Cauchy slice to $\mathscr{I}$:
\begin{equation} \label{ASv1}
 \boldsymbol{\Omega}_{\mathscr{I}} =\frac1{8\pi G} \int_{\mathscr{I}} dv d^2 x \sqrt{q} \delta \lambda_1 \wedge \delta \sigma_0 +c.c.
\end{equation} where $c.c.$ denotes the complex conjugate terms. The corresponding Poisson bracket is used to compute the $Lw_{1+\infty}$ charge algebra.

\end{itemize}

As we have explained in Section \ref{sec:Covariantization}, the recursion relations \eqref{full recursion at null infinity} also exist at the horizon (see Equation \eqref{Bianchi evolution equations}) and require the introduction of Weyl covariant operators. This amounts to substitute $\eth \to \eth_{\mathscr{C}}$ and $-\partial_v \to \pthorn'_{\mathscr{C}}$ in the above relation. Furthermore, because of their Weyl-covariant form, the recursion relations can be directly mapped from null infinity to the horizon via conformal compactification. In Section \ref{sec:canonical symmetries at the horizon}, we shall explain that, upon self-duality
assumptions, the analogue of the Ashtekar-Streubel symplectic structure \eqref{ASv1} will appear at
the subleading order in $r$ at the horizon. Although the difference between the full and self-dual theory of gravity only appears in
the subleading orders at null infinity (for spin $s \ge 4$), it appears at leading order at the
horizon. This key difference makes the whole analysis much more intricate at finite distance
than at null infinity. More precisely, as we shall see in the next section, the assumption of
self-duality will constrain the transverse dyad induced at the horizon, which is leading in
the radial expansion.

\subsection{Subleading tower of surface charges}
\label{sec:Subleading tower of surface charges}

Using a heuristic construction mimicking the approach used at null infinity \cite{Freidel:2021dfs,Freidel:2021ytz,Geiller:2024bgf}, we now write the canonical generator of the $Lw_{1+\infty}$ symmetries at the horizon. The explicit expressions for the charges are known at null infinity, for any cut $v = \text{constant}$ of $\mathscr{I}$. These charges are built to satisfy two properties: $(i)$ they are conserved at $\mathscr{I}$ in absence of radiation, characterized by $\Psi^0_4 = Q_{-2} = 0$ (see also \cite{Kmec:2024nmu,Cresto:2024mne}), and $(ii)$ their associated integrated fluxes satisfy the $Lw_{1+\infty}$ algebra using the Ashtekar-Streubel bracket. Let us attempt to extend and adapt this construction at the horizon while maintaining the same criteria. The lowest spin charge is given by
\begin{equation}
    H_{-1} = \frac{r}{8\pi G} \oint T_{-1} Q_{-1} 
\label{spinm1charge}
\end{equation} where $T_{-1}$ is the symmetry parameter associated with the spin $-1$ symmetry and the integral
    \begin{equation}
    \oint =\int d^2x \sq
\end{equation} is performed over a $2$-surface $v= \text{constant}$ and $r= \text{constant}$ close to the horizon. The  global factor $r$ will be justified in Section \ref{sec:canonical symmetries at the horizon} and follows from the fact that the charges constructed here will be associated with the subleading phase space at the horizon. The flux associated with \eqref{spinm1charge} is given by 
\begin{equation}
    -\partial_v H_{-1}= \frac{r}{8\pi G}\oint \pthorn'_{\mathscr{C}} ( T_{-1} Q_{-1}  )   =\frac{r}{8\pi G}\oint( \pthorn'_{\mathscr{C}} T_{-1} Q_{-1} + T_{-1} \eth_{\mathscr C}Q_{-2}) :=F_{-1}
\end{equation} where, in the first equality, we used 
\begin{equation}
    \pthorn'_{\mathscr{C}} \sqrt{q} = 0
\end{equation} by definition of $\pthorn'_{\mathscr{C}}$ in \eqref{thorn prime} and $\mu_0 = \frac{1}{2}\partial_v\ln \sqrt{q}$, and in the second equality, we used the recursion relation \eqref{Bianchi evolution equations} for $s= -1$. Furthermore, we assume $
    \pthorn'_{\mathscr{C}} T_{-1} = 0$, which generalizes the condition $\partial_v T_{-1} = 0$ at null infinity to any null hypersurface. At the end, we have
\begin{equation}
    F_{-1} = \frac{r}{8\pi G}\oint( T_{-1} \eth_{\mathscr C}Q_{-2})
\end{equation} This flux is formally the same as the one considered at $\mathscr{I}$ \cite{Kmec:2024nmu}, and vanishes as $Q_{-2} = 0$. 

Let us now focus on the case $s= 0$. Again, inspired by the charge expression at null infinity, we start from the ansatz
\begin{align}
    H_{0}&=\frac{r}{8\pi G}\oint T_0\left( Q_0 - U \, \eth_{\mathscr C}Q_{-1}\right)
\end{align}
with $U:\{1,1,0\}$ a `dressed time coordinate' satisfying $\pthorn'_{\mathscr C} U=1$, generalizing the simple $v$ coordinate satisfying $\partial_v v = 1$. An explicit solution is given by 
\begin{equation}
    U=e^{\int^v dv' \left( \gamma_0+\bar\gamma_0  +2 \mu_0\right)}\left(C(x^A)-\int^v dv'e^{-\int^{v'} dv'' \left( \gamma_0+\bar\gamma_0  +2 \mu_0\right)} \right)\,. \label{solution U}
\end{equation} where $C(x^A)$ will be fixed later on.  As in the previous case, we will also assume $\pthorn_{\mathscr{C}}' T_{0} = 0$. Without any further assumption we have 
\begin{equation}
\begin{split}
-\partial_vH_{0}&=\frac{r}{8\pi G}\oint T_0 \big( -\sigma_0 \,Q_{-2}-U \, \pthorn'_{\mathscr C}(\eth_{\mathscr C}Q_{-1})\big)\\ 
&=\frac{r}{8\pi G}\oint T_0 \left( -\sigma_0 \,Q_{-2}-U \,\eth^2_{\mathscr C}Q_{-2}\right) \\\label{F0secondline}
&-T_0\,U  \left( \bar\lambda_0  \eth'_{\mathscr C} Q_{-1}
+2Q_{-1}\Big[\partial_0\mu_0+\mu_0\tau_0+(\bar\partial_0+3\alpha_0-\bar\beta_0)\bar\lambda_0+\partial_0(\gamma_0+\bar\gamma_0)
\Big]  \right) .
\end{split}
\end{equation} To obtain the last line, we used the commutation relation displayed in \eqref{commutation thornprimeEth}. The last line does not vanish when $Q_{-2}=0$, hence violating the expected conservation law. One could try to modify the expression for $H_0$ to absorb these extra terms and produce only contributions with $Q_{-2}$. However, the fact that the commutator between $\eth'_{\mathscr C}$ and $\pthorn'_{\mathscr C}$ is non zero but given by \eqref{commbareththornprim} will bring new terms that do not cancel out and produce more terms involving spin coefficients.\footnote{
The second line of \eqref{F0secondline} is non vanishing when imposing $Q_{-2}=0$.  We can try to absorb that term by shifting the charge $H_0$. For instance
\begin{equation}
    \tilde H_0=H_0+\frac{r}{8\pi G}\oint T_0 \frac{U^2}2\bar\lambda_0  \eth'_{\mathscr C} Q_{-1}
\end{equation}
which evolves as
\begin{equation}
-\partial_v \tilde H_0=-\partial_vH_0+\frac{r}{8\pi G}\oint T_0 \left( U \bar \lambda_0  \eth'_{\mathscr C} Q_{-1}+ \frac{U^2}2 \pthorn'_{\mathscr C} \bar \lambda_0 \eth'_{\mathscr C} Q_{-1}+ \frac{U^2}2 \bar \lambda_0 \pthorn'_{\mathscr C} \eth'_{\mathscr C} Q_{-1}\right)
\end{equation}
The first term will cancel the first term in \eqref{F0secondline}, the second term is proportional to $\bar Q_{-2}$ when using \eqref{evolambda}. However for the third term, we need to use \eqref{commbareththornprim} which adds a bunch of terms that will not vanish or cancel against the last term in \eqref{F0secondline}.} Another option would be to impose by hand $Q_{-1}=0$, however similar issues will happen for higher spin charges. After trials and errors, this turns out to be impossible, and these issues will even get worse for higher spin charges.

Instead, we propose to work with a restricted solution space. Two choices are interesting to consider: 
\begin{subequations}
\begin{align}\label{dualcond1}
(i) \quad \bar\lambda_0&=0\,,\quad \bar{\gamma}_0=0\,,\quad \gamma_0 =-\kappa\,, \quad \mu_0=0\\
(ii) \quad \bar\lambda_0&=0\,,\quad \bar{\gamma}_0=0\,,\quad \gamma_0=-\kappa\,,\quad \mu_0=\kappa\,,\quad\tau_0=0 \,.\label{dualcond1prime}
\end{align}
\end{subequations}
Here $\kappa$ is now constant, which is the case for a black hole or a cosmological horizon. We work in a complexified spacetime. For instance $\bar{\lambda}_0$ is considered as a complex function independent from $\lambda_0$, so that, in this restricted solution space, $\lambda_0$ is generically non zero. Both conditions \eqref{dualcond1} and \eqref{dualcond1prime} ensure $[\pthorn'_{\mathscr C},\eth_{\mathscr C}]=0$ (see Equation \eqref{commbareththornprim}) so that the second term of \eqref{F0secondline} vanishes. Note that $[\pthorn'_{\mathscr C},\eth'_{\mathscr C}]\neq0$ even after imposing these conditions. These conditions can be interpreted as self-duality conditions, fixing one of the two helicities on the phase space. Identifying self-duality conditions at finite distance is a delicate questions. On the one hand, the condition \eqref{dualcond1} was used in \cite{Ruzziconi:2025fct} for the discussion of celestial symmetries on null hypersurfaces. On the other hand, black holes and horizons in self-dual gravity have been discussed e.g. in \cite{Crawley:2021auj,Crawley:2023brz,Guevara:2023wlr,Giribet:2025ihk}. In Section \ref{sec:Self-dual Kleinian Taub-NUT black holes}, we revisit the self-dual Kleinian Taub-NUT black hole and show that it is included in the solution space defined by \eqref{dualcond1prime}. It is instructive to verify that both conditions \eqref{dualcond1} and \eqref{dualcond1prime} are consistent with the evolution equations displayed in Appendix \ref{sec:Relevant equations} and constrain only one helicity sector, hence allowing for non-vanishing transverse radiation. In the rest of this section, for concreteness, we will focus on condition \eqref{dualcond1}, but a completely analogue discussion would hold for \eqref{dualcond1prime}. In this restricted phase space, we have 
\begin{equation}
    F_0=\frac{r}{8\pi G}\oint T_0 \left( -\sigma_0 \,Q_{-2}-U \,\eth^2_{\mathscr C}Q_{-2}\right) 
\end{equation} which vanishes when $Q_{-2} = 0$, as desired.

Assuming \eqref{dualcond1}, we can easily proceed with the higher spin charges starting from the expressions at null infinity, and performing the substitutions $\partial_v \to -\pthorn'_{\mathscr{C}}$, $\eth \to \eth_{\mathscr{C}}$, $v \to - U$, and with parameters $T_s(u,x^A)$ satisfying
\begin{equation}
    \boxed{\pthorn'_{\mathscr{C}} T_s = 0, \qquad s= -1, 0, 1, 2 , \ldots }
\end{equation} Here $T_s$ has weights $\{-2s;0;1+s\}$, so that the integrand in $H_s$ has weights $\{0;0;0\}$. Notice that the solution \eqref{solution U} reduces to
$U=-\frac1\kappa+ C(x^A) e^{-\kappa\,v}$. We choose the constant $C=1/\kappa$ such that for $\kappa\to0$, $U$ reduces to the undressed time $v$. This leads to
\begin{equation}
\boxed{U=\frac1\kappa \left(-1+e^{-\kappa v}\right) = -v+\mathcal O(\kappa). }   
\end{equation}
We display below the explicit expressions of the surface charges up to spin $2$: 
\begin{equation}
   \boxed{ \begin{split}
        \label{surface charges v constant}
H_{-1}&=\frac{r}{8\pi G}\oint T_{-1}Q_{-1} ,\\
H_{0}&=\frac{r}{8\pi G}\oint T_0\left( Q_0 -U\, \eth_{\mathscr C}Q_{-1}\right) ,\\
H_{1}&=\frac{r}{8\pi G}\oint T_1 \left[Q_1 -U \eth_{\mathscr C} Q_{0}+\frac{U^2}2\eth_{\mathscr C}^2Q_{-1}+2Q_{-1}(\pthorn_{\mathscr C}')^{-1}\sigma_0\right] ,\\
H_{2}&=\frac{r}{8\pi G}\oint T_2 \Big[Q_2 -U \eth_{\mathscr C} Q_{1}+\frac{U^2}2 \eth_{\mathscr C}^2Q_0-\frac{U^3}6 \eth_{\mathscr C}^3 Q_{-1}\\
&\qquad\qquad+3Q_0(\pthorn_{\mathscr C}')^{-1}\sigma_0 -3\eth_{\mathscr C}Q_{-1}(\pthorn_{\mathscr C}')^{-2}\sigma_0-2\eth_{\mathscr C}(Q_{-1}(\pthorn_{\mathscr C}')^{-1}(U\sigma_0))\Big] .
 \end{split} }
\end{equation}  The associated fluxes $F_s = - \partial_v H_s$ can be computed using \eqref{Bianchi evolution equations} and read as
\begin{equation}  \label{fluxes cut}
\begin{split}
F_{-1}&=\frac{r}{8\pi G}\oint T_{-1}\eth_{\mathscr C} Q_{-2} , \\
F_{0}& =\frac{r}{8\pi G}\oint T_0 \left( -\sigma_0 \,Q_{-2}-U \,\eth^2_{\mathscr C}Q_{-2}\right) ,\\
F_{1}& =\frac{r}{8\pi G}\oint T_1 \Big[U\eth_{\mathscr C}(\sigma_0Q_{-2}) +\frac{U^2}2 \eth^3_{\mathscr C}Q_{-2}+2\eth_{\mathscr C}Q_{-2}(\pthorn_{\mathscr C}')^{-1}\sigma_0\Big] , \\
F_{2}&=\frac{r}{8\pi G}\oint T_2 \Big[-\frac{U^2}2\eth_{\mathscr C}^2(\sigma_0Q_{-2})-\frac{U^3}6\eth_{\mathscr C}^4Q_{-2}
-3\sigma_0Q_{-2}(\pthorn_{\mathscr C}')^{-1}\sigma_0
-3\eth_{\mathscr C}^2Q_{-2} (\pthorn_{\mathscr C}')^{-2}\sigma_0\\
&\qquad\qquad\qquad-2\eth_{\mathscr C}(\eth_{\mathscr C}Q_{-2}(\pthorn_{\mathscr C}')^{-1}(U\sigma_0))\Big] .
\end{split}
\end{equation}
 Those vanish consistently when $Q_{-2} = 0$, providing a tower of conserved charges at the black hole horizon, in absence of transverse radiation. This also clearly identifies $Q_{-2}$ as the transverse radiation through the horizon, responsible for the non-conservation. As we shall explain in the next section, the integrated fluxes over the horizon, $\int_{-\infty}^{+\infty} dv F_s$, coincide with the canonical generators of the $Lw_{1+\infty}$ symmetries at the horizon for the subleading symplectic structure, hence satisfying the exact same criteria than those required at null infinity. Let us emphasize that the charges \eqref{surface charges v constant} and their associated fluxes \eqref{fluxes cut} provide useful dynamical information for the solution at finite distance through the flux-balance laws.

\subsection{Canonical $Lw_{1+\infty}$ symmetries at the horizon}
\label{sec:canonical symmetries at the horizon}

As already suggested in the previous sections, we will show that the analogue of the Ashtekar-Streubel symplectic structure will appear at subleading order in the $r$ expansion. First, we show that under the self-duality assumption \eqref{dualcond1} and  additional consistent conditions on the variation of phase space \eqref{dualcond2}, the leading symplectic structure vanishes.

We consider again the radial expansion of the presymplectic potential \eqref{expansion Theta}. In NP formalism, the leading order \eqref{leading presymplectic pot} can be rewritten as
\begin{equation}
    \Theta^r_{(0)} =-\frac{\sq}{8\pi G}\left( \bar\lambda_0 \,\bar m^A_0\delta \bar m_A^0+\lambda_0\,m^A_0 \delta  m_A^0 +\delta \left(\mu_0-(\gamma_0+\bar\gamma_0)\right) \right) +  \delta\left(\frac{\sq}{8\pi G} \mu_0\right) . 
    \label{leading presumpelcticNP}
\end{equation} To obtain this expression, let us emphasize that we have just imposed the radial expansions discussed in Sections \ref{sec:Phase space in metric formalism} and \ref{sec:From metric to Newman-Penrose formalism} (i.e. we have used \eqref{MetricTaylorexpansion} and \eqref{solspace} without imposing the time evolution equations displayed in Appendix \ref{sec:Relevant equations}). Remarkably, the expression \eqref{leading presumpelcticNP} vanishes when imposing \eqref{dualcond1}, together with
\begin{equation}
  \delta m_A^0=0 \,,\quad \delta \kappa=0  \,, \quad\delta \sqrt{q}=0 . \label{dualcond2}
\end{equation}
Again, imposing these conditions is consistent with the equations of motion \eqref{evolmbA}. Using $\delta\ln \sqrt{q}=m^A_0\delta\bar m_A^0+\bar m^A_0\delta m_A^0$, \eqref{dualcond2} also implies $\delta m^A_0=0$. Hence, the leading order phase space completely trivializes and does not seem to play any role in the identification of the celestial symmetries in the self-dual sector. In addition, in the following, it will also be convenient to impose 
\begin{equation}
    \delta\tau_0 = 0
    \label{dualcond3}
\end{equation} yielding $\delta \bar{\alpha} = \delta \beta = 0$. This implies that variations on the phase space commute with derivative operators appearing in the charges derived in Section \ref{sec:Subleading tower of surface charges}, i.e.
\begin{equation}
    [\delta , \eth_{\mathscr{C}}] = 0, \qquad [\delta, \pthorn'_{\mathscr{C}}] = 0 . 
    \label{commutation useful}
\end{equation} The solution space resulting from the self-duality conditions \eqref{dualcond1}, \eqref{dualcond2} and \eqref{dualcond3} is fully consistent with respect to the equations in Appendix \ref{sec:Relevant equations} and still generically radiative (with one helicity) since $\lambda_0 \neq 0$.

Next, taking the radial expansions \eqref{MetricTaylorexpansion} and \eqref{solspace} and the self-duality conditions \eqref{dualcond1} and \eqref{dualcond2} into account, the subleading presymplectic potential is 
\begin{equation}\label{subleading symplt pot}  
    \Theta^r_{(1)}=\frac{\sqrt{q}}{8\pi G} (-\lambda_0 \delta\sigma_0 + \left(( \partial_v-2\lambda_0+2\kappa)\sigma_0 +2(\eth\tau_0 +\tau_0^2)\right)\,\bar m_A^0\delta \bar m^A_0)+\delta \mu^r+\partial_vY^{vr}+\partial_AY^{Ar} . 
\end{equation}
The three last terms can be eliminated by Iyer-Wald ambiguities \cite{Iyer:1994ys}, and we will discard them. The expression of the symplectic potential without the self duality conditions \eqref{dualcond1} and \eqref{dualcond2} at subleading order, including ambiguities, was given in metric formalism in Section \ref{sec:sympletic pot}, see Equation \eqref{subleading sympl pot metric}.

Taking the above conditions into account, the symplectic form at the horizon is subleading and is given by 
\begin{equation}
    \boldsymbol{\Omega}_\mathcal{H} = \int_{r = 0} \delta \Theta^r (d^3 x)_r 
 =  \frac{r}{8\pi G}\oint \int dv \,\delta \sigma_0 \wedge \delta \lambda_0 + \delta \bar{p}^A \wedge \delta \bar{m}^0_A +\mathcal{O}(r^2)
\label{subleading symplectic0}
\end{equation} where $\bar p^A =-\left(( \partial_v-2\lambda_0+2\kappa)\sigma_0 +2(\eth\tau_0 +\tau_0^2)\right)\,\bar m^A_0$. A crucial observation is that $\bar m_A$ does not appear in the flux expressions \ref{fluxes cut} (at this stage, $\bar m_A$ and $\lambda_0$ are treated as independent canonical variables --- they will be related once we are fully on-shell, after imposing $\lambda_0 =\bar m^{A}_0(\partial_v +  \gamma_0)\bar m_A^0$, see Equation \eqref{evolmbA}). Therefore, the second symplectic pair will not play any role in the computation of the flux algebra. For this reason, we will focus on the first symplectic pair $(\lambda_0, \sigma_0)$, as it is also done in treatment of celestial symmetries at null infinity \cite{Freidel:2021ytz,Geiller:2024bgf, Kmec:2024nmu}.\footnote{The discussion of $Lw_{1+\infty}$ symmetries at null infinity \cite{Freidel:2021ytz,Geiller:2024bgf, Kmec:2024nmu}, which includes superrotation symmetries corresponding to $s=1$, would in principle require enhancing the phase by allowing fluctuation of the boundary metric \cite{Campiglia:2014yka,Campiglia:2015yka,Compere:2018ylh}. This would yield terms similar to the second term in \eqref{subleading symplectic0}. However, these terms are discarded in that context and the symplectic structure reduces to Ashtekar-Streubel. 
} Hence, we are left with the following symplectic structure at the horizon: 
\begin{equation}
   \boxed{ \boldsymbol{\Omega}_{\mathcal{H}} = \frac{r}{8\pi G} \oint \int dv\, \delta \sigma_0 \wedge \delta \lambda_0   +\mathcal{O}(r^2)}
\label{subleading symplectic}
\end{equation} which is the analogue of the Ashtekar-Streubel symplectic structure \eqref{ASv1} at finite distance. Hence, we have demonstrated that this symplectic structure naturally appears at subleading order at the horizon. By contrast with null infinity, $\lambda_0$ and $\sigma_0$ are no longer related when imposing all the equations of motion at finite distance: $\lambda_0$ is the intrinsic shear, while $\sigma_0$ is the extrinsic shear. Notice that $\delta \sqrt{q}= 0$ (see \eqref{dualcond2}), which is analogous to null infinity: $(\sigma_0, \lambda_0)$ form a canonical pair and satisfy the following bracket: 
\begin{equation}
    \boxed{\{ \sigma_0 (v_1, x^A_1) , \lambda_0 (v_2, x^A_2) \} = \frac{8 \pi G}{r \sqrt{q}}  \delta^2 (x^A_1 - x^A_2) \delta (v_1-v_2) . }
    \label{canonical bracket}
\end{equation} 

We can now use this bracket to show that the charges constructed heuristically in Section \ref{sec:Subleading tower of surface charges} are the canonical generators for the $Lw_{1+\infty}$ symmetries at the horizon. The spin $s=-1,0,1,2$ integrated fluxes are defined by $\mathcal{F}_s =\int dv F_s$, where $F_s$ are given in \eqref{fluxes cut}.  We have explicitly
\begin{equation} \label{integrated flux expressions} \boxed{
\begin{split} 
 \mathcal F_{-1}&=\frac r{8\pi G} \int dv \oint T_{-1}\eth_{\mathscr C} Q_{-2} ,\\
 \mathcal F_{0}& =\frac r{8\pi G}\int dv  \oint T_0 \left[ -\sigma_0 \,Q_{-2}-U \,\eth^2_{\mathscr C}Q_{-2}\right] ,\\
 \mathcal F_{1}& =\frac r{8\pi G} \int dv \oint T_1 \Big[U\eth_{\mathscr C}(\sigma_0Q_{-2}) +\frac{U^2}2 \eth^3_{\mathscr C}Q_{-2}+2\eth_{\mathscr C}Q_{-2}(\pthorn_{\mathscr C}')^{-1}\sigma_0\Big] , \\
 \mathcal F_{2}&=\frac r{8\pi G} \int dv \oint T_2 \Big[-\frac{U^2}2\eth_{\mathscr C}^2(\sigma_0Q_{-2})-\frac{U^3}6\eth_{\mathscr C}^4Q_{-2}
-3\sigma_0Q_{-2}(\pthorn_{\mathscr C}')^{-1}\sigma_0\\
&\qquad\qquad\qquad\qquad\qquad-3\eth_{\mathscr C}^2Q_{-2} (\pthorn_{\mathscr C}')^{-2}\sigma_0-2\eth_{\mathscr C}(\eth_{\mathscr C}Q_{-2}(\pthorn_{\mathscr C}')^{-1}(U\sigma_0))\Big]
\end{split} }
\end{equation} 
Notice that $\mathcal{F}_s$ is finite only if fields decrease sufficiently fast at the horizon when $v \to \pm \infty$. To avoid any of these potential divergences, similarly to what is done at null infinity \cite{Freidel:2021ytz, Geiller:2024bgf, Freidel:2022skz, Kmec:2024nmu}, we will impose Schwartzian falloffs on the fields, i.e. $\lim_{v \to \pm \infty}\lambda_0 \sim e^{-|v|^2}$. In practice, this will allow us to neglect total derivative terms with respect to $v$ inside of the flux integrals. 

Using the integrated flux expressions \eqref{integrated flux expressions} and the canonical bracket \eqref{canonical bracket}, we deduce the transformation of $\sigma_0$ from 
\begin{equation}
    \delta_s \sigma_0 = \{ \mathcal{F}_s , \sigma_0 \}
\end{equation} 
To compute this, it is easier to make some integration by parts before evaluating the bracket. For instance, $\mathcal F_{0}=\frac r{8\pi G}\int dv  \oint (-\pthorn_{\mathscr C}') \left[ T_0\sigma_0+\eth^2_{\mathscr C}(T_0U) \right] \lambda_0$.
Explicitly, we find
\begin{subequations}
    \begin{align}
        \delta_{-1}\sigma_0&=0\\\label{delta0sigma}
        \delta_{0}\sigma_0&
        = T_0\pthorn_{\mathscr C}'\sigma_0 +\eth^2_{\mathscr C}T_0
        \\\label{delta1sigma}
        \delta_{1}\sigma_0& 
        =\left(3 \eth_{\mathscr C}T_1+2T_1\eth_{\mathscr C}+U\eth_{\mathscr C}T_1\pthorn_{\mathscr C}'\right)\sigma_0+U\eth^3_{\mathscr C}T_1\\ \nonumber
        \delta_{2}\sigma_0&=\frac{U^2}2\left(\eth^4_{\mathscr C} T_2-\eth^2_{\mathscr C}T_2 \pthorn'_{\mathscr C}\sigma_0 \right) +U\left(3\eth^2_{\mathscr C}T_2\sigma_0+2\eth_{\mathscr C} T_2\eth_{\mathscr C}\sigma_0\right)\\&+3\left(T_2 \pthorn_{\mathscr C}'\sigma_0+\eth^2_{\mathscr C}T_2\right)(\pthorn_{\mathscr C}')^{-1}\sigma_0
    +3T_2\sigma_0^2
    \end{align}
\end{subequations}
consistently with the transformations found at null infinity \cite{Kmec:2024nmu}. Similarly, we find
\begin{subequations}
    \begin{align}
        \delta_{-1}\lambda_0&=0\\\label{delta0lambda}
        \delta_{0}\lambda_0&=-T_0Q_{-2}
        \\\label{delta1lambda}
        \delta_{1}\lambda_0&= -U\eth_{\mathscr C}T_1 Q_{-2}+2 T_1\eth_{\mathscr C}\lambda_0 
       %\\ \delta_2\lambda_0&= -\frac{U^2}2\eth^2_{\mathscr C}T_2Q_{-2}+3\eth^2_{\mathscr C}(\pthorn_{\mathscr C}')^{-1}\lambda_0-2 U\eth_{\mathscr C}T_2 \eth_{\mathscr C}\lambda_0 + \int -3Q_{-2} \{\sigma_0(\pthorn_{\mathscr C}')^{-1}\sigma_0,\lambda_0\}
    \end{align}
\end{subequations}

Finally, using again \eqref{canonical bracket} and \eqref{integrated flux expressions}, and imposing the self-duality conditions \eqref{dualcond1}, \eqref{dualcond2}, \eqref{dualcond3}, so that \eqref{commutation useful} holds, the computation of the algebra is the exact analogue of the one performed at null infinity \cite{Freidel:2021ytz,Geiller:2024bgf} (see also \cite{Cresto:2024fhd,Cresto:2024mne} for the non-linear computation). In summary, we find 
\begin{equation}
   \boxed{\{ \mathcal{F}_{T_{s_1}}, \mathcal{F}_{T_{s_2}} \} = \mathcal{F}_{T_{s_1+ s_2 - 1}}, \qquad T_{s_1+ s_2 - 1} = ({s_2} + 1) T_{s_2}\eth_{\mathscr{C}} T_{s_1} - ({s_1} + 1)T_{s_1} \eth_{\mathscr{C}} T_{s_2} }
     \label{representation of Linfinity}
\end{equation} where we noted $\mathcal{ F}_s \equiv \mathcal{F}_{T_s}$ (at a non-linear-level, this bracket becomes field-dependent and the whole structure becomes a Lie algebroid \cite{Cresto:2024fhd,Cresto:2024mne}). The symmetry parameters are sometimes required to satisfy the wedge condition $
\eth_{\mathscr{C}}^{s+2} T_s = 0$. With this condition, the algebra corresponds to the wedge subalgebra of $Lw_{1+\infty}$. Hence, we have shown that $\mathcal{F}_s$ are the canonical generators for the celestial symmetries at the horizon. 

We conclude this section with some comments. At this stage, one may wonder whether the parameters $T_s$ have a spacetime interpretation. Analogously to what happens at null infinity \cite{Freidel:2021dfs,Freidel:2021ytz,Geiller:2024bgf}, only the spin $s=0,1$ symmetries have a diffeomorphism interpretation. In Section \ref{app:charges} we derived the residual gauge transformations for the solution space discussed in Section \ref{sec:Phase space in metric formalism}. The relation with spin $0$ and spin $1$ symmetries is discussed in Appendix \ref{sec:Diffeomorphism interpretation}. Furthermore, the subleading charges computed in Section \ref{sec:Phase space in metric formalism} exhibit patterns of the spin $0$ and spin $1$ charges construed \eqref{surface charges v constant}. However, the higher spin charges do not have such an interpretation on spacetime. As it is the case at null infinity \cite{Kmec:2024nmu}, we expect that a twistor space formulation of self-dual gravity adapted to finite-distance null hypersurfaces will provide such an interpretation. 

In the above discussion, we considered explicit expressions up to spin $s= 2$. This is in principle sufficient to generate the higher spin symmetry generators $\mathcal{ F}_s$, $s>2$, by just using the bracket \eqref{representation of Linfinity} recursively. Again, these expressions will be the formally the same as those derived at null infinity, up to the above mentioned substitutions. We refer to \cite{Cresto:2024fhd, Cresto:2024mne} for non-perturbative spacetime expressions. To obtain the surface charges $H_s$ on a $2$-surface near the horizon, we would need to identify the $Q_s$ near the horizon for $s>2$. As explained in Appendix \eqref{sec:Radial expansion}, the structure of the radial expansion is slightly different from null infinity, but one can still extract some $Q_s$, for $s> 2$, from the expansion of $\Psi_0$. At null infinity, a self-dual truncation of the Bianchi identities leads to the recursion relations \eqref{Bianchi evolution equations} for all $s>-2$. We expect a similar feature here and leave this discussion for future endeavour.

\subsection{Self-dual Kleinian Taub-NUT black holes}
\label{sec:Self-dual Kleinian Taub-NUT black holes}

In this last section, we show that the self-dual Kleinian Taub-NUT black hole solutions are part of the phase space defined by the conditions \eqref{dualcond1prime}.

Starting from the Taub-NUT solution parametrized by the mass $M$ and the NUT charge $N$, the self-duality condition on the metric in Kleinian $(2,2)$ signature implies $M=N$. The self-dual Kleinian Taub-NUT metric is then parametrized by $M$ and is given by 
\begin{equation}
 ds^2= \frac{\tilde r-M}{ \tilde r+M}(d\tilde t-2M\cosh\tilde \theta d \tilde \phi)^2+ \frac{\tilde r+M}{\tilde r-M}d\tilde r^2+(M^2-\tilde r^2)(d\tilde \theta^2+\sinh^2\tilde \theta  d\tilde \phi^2) 
\end{equation}
with $\tilde r\in[-M,\infty)$, $\tilde \theta\in[0,+\infty)$ and $\tilde t,\tilde \phi$ obey the periocity conditions $(\tilde t,\tilde \phi)\sim (\tilde t+4\pi M,\tilde \phi+2\pi)$. We refer to \cite{Crawley:2021auj} for more details, and \cite{Crawley:2023brz,Bogna:2024gnt,Heuveline:2025nmb,Adamo:2025fqt} for related discussions in the context of celestial symmetries. 

We perform a change of coordinates to reach the near-horizon coordinates. For that we use the advanced time $v$ via $\tilde \theta=\frac{v}{2M}-\log\frac{\rho}{2M}$ and $\rho=\tilde r-M$ \cite{Giribet:2025ihk}. We further reach the coordinates \eqref{metricGNC} via $\rho =2 \sqrt{ M^2+ M r}\,-2 M$, $\tilde t=x^1 \,M$, $\tilde \phi= x^2$ (we have the periodicity $(x^1,x^2)\sim(x^1+4\pi,x^2+2\pi)$). We obtain 
\begin{align}
V&= \frac{r+M-\sqrt{M (M+r)}}{M} \,, \qquad U^A= 0 , \nonumber\\
\gamma_{12}&=-\frac{M^2 e^{-\frac{v}{2 M}}}{\sqrt{M (M+r)}} \left(r-2 \sqrt{M (M+r)}+M \left(e^{v/M}+2\right)\right) , \\
\gamma_{11}&= M^2-\frac{M^3}{\sqrt{M (M+r)}} , \nonumber\\\nonumber
\gamma_{22}&=
\frac{e^{-\frac{v}{M}} \left(r-M e^{v/M}\right)}{\sqrt{M (M+r)}} \left(M^2 \left(e^{v/M}+4\right)+M \sqrt{M (M+r)} \left(e^{v/M}-4\right)+3 M r-r \sqrt{M (M+r)}\right) . 
% \frac{4 M \,r \, e^{-\frac{v}{M}}-1}{4 \sqrt{M (M+r)}} \times\\
%-  \frac{e^{-\frac{v}{M}} \left(e^{\frac{v}{M}}-4 M r\right)}{4 \sqrt{M (M+r)}} \times \nonumber\\
%&\times\left(4 M \left(4 M^2+3 M r-4 M \sqrt{M (M+r)}-r \sqrt{M (M+r)}\right)+\left(\sqrt{M (M+r)}+M\right) e^{\frac{v}{M}}\right)\,. \nonumber
\end{align}
The radial expansion near the horizon at $r=0$ is given by
\begin{align}
V&= \frac{r}{2 M}+\frac{r^2}{8 M^2}+O\left(r^3\right) ,\nonumber\\
\gamma_{12}&=-M^2 e^{\frac{v}{2 M}}+\frac{r}{2} M e^{\frac{v}{2 M}}-\frac{r^2}8e^{-\frac{v}{2 M}} \left(3 e^{v/M}+2\right)+O\left(r^3\right) ,\\
\gamma_{11}&=\frac{r}{2 }M-\frac{3 r^2}{8}+O\left(r^3\right) ,\nonumber\\
\gamma_{22}&=-2 M^2 e^{v/M}+\frac{1}{2} M r \left(e^{v/M}+4\right)-\frac{1}{8} r^2 \left(3 e^{v/M}+4\right)+O\left(r^3\right) . \nonumber
\end{align}
In particular we have
\begin{equation}
    q_{AB}=-2M^2\left(e^{\frac {v}{2M}}dx^1dx^2+e^{\frac{v}M}(dx^2)^2\right) . 
\end{equation}
The dyad \eqref{transverse dyad} $m_A^0$ is given by $m_A^0=-M e^{\frac{v}{2 M}}dx^2$ and $\bar m_A^0=M(dx^1+e^{\frac{v}{2 M}}dx^2)$. Note that in Klein signature, $m_A^0$ and $\bar m_0^A$ are real and independent elements, i.e. they are not complex conjugate of each other. This dyad does not satisfy the conditions \eqref{dualcond2}. However this is easily solved by using the freedom we have in the tetrad to rescale $m_A^0\to -A\, m_A^0$, $\bar m_A^0\to -\frac1A \bar m_A^0$. We take $A=\frac{e^{-\frac{v}{2 M}}}{M}$. The new dyad is 
\begin{equation}
    m_A^0= dx^2\,,\quad \bar m_A^0=-M^2e^{\frac{v}{2 M}}(dx^1+e^{\frac{v}{ M}}dx^2)
\end{equation}
which now satisfies $\delta m_A^0=0$. 

Due to the change of signature, we recompute the spin coefficients in the NU tetrad. We have
\begin{equation}  \alpha=\bar\alpha=\beta=\bar\beta=\tau=\bar\tau=\nu=\bar\nu=\bar\gamma=O\left(r^3\right)
\end{equation}
and the non-vanishing spin-coefficients are 
\begin{align}
\mu&=\bar\mu=\frac{1}{4 M}+\frac{r^2 e^{-\frac{v}{M}}}{4 M}+O\left(r^3\right)\,,\qquad \rho=\bar\rho=-\frac{r e^{-\frac{v}{M}}}{4 M^2}+\frac{3 r^2 e^{-\frac{v}{M}}}{16 M^3}+O\left(r^3\right) , \nonumber \\
  \gamma&=-\frac{1}{4 M}-\frac{r}{8 M^2}+\frac{3 r^2}{32 M^3}+O\left(r^3\right) , \\
\sigma&=-\frac{e^{-\frac{v}{M}}}{4 M^3}+\frac{3 r e^{-\frac{v}{M}}}{8 M^4}-\frac{r^2 e^{-\frac{2 v}{M}} \left(15 e^{v/M}+2\right)}{32 M^5}+O\left(r^3\right) \,,\quad 
\bar\sigma=-M-\frac{r^2 e^{-\frac{v}{M}}}{4 M}+O\left(r^3\right) ,  \nonumber \\
%%%%%%%%%%%
  \lambda&=\frac{M}{2} e^{v/M}+\frac{r}{4}+\frac{r^2}{16 M}+O\left(r^3\right) \,,\quad \bar\lambda=\frac{r e^{-\frac{v}{M}}}{16 M^4}-\frac{3 r^2 e^{-\frac{v}{M}}}{64 M^5}+O\left(r\right)^3 . \nonumber
\end{align} These are compatible with the self-duality conditions \eqref{dualcond1prime}. 
The remaining conditions in \eqref{dualcond2} (we already have achieved $\delta m_A^0=0$) are enforced by imposing $\delta M=0$. For this reason, it is important to restrict the phase space to the particular solution considered here only after performing all the variations.

The Weyl scalars are 
\begin{equation}
\begin{split} \label{sdWeyl}
    \Psi_4^0&=\frac{3 e^{v/M}}{8} \,,\quad \Psi^0_2 =-\frac{1}{8 M^2}\,,\quad \Psi_0^0=\frac{3 e^{-\frac{v}{M}}}{8 M^4}\,, \quad \Psi_0^1=-\frac{15  e^{-\frac{v}{M}}}{16 M^5} , \\
    \Psi^0_3&=\Psi_1^0=0 , \\
    \bar\Psi_4^0&=\bar\Psi_3^0=\bar\Psi_2^0=\bar\Psi_1^0=\bar\Psi_0^0=0 . 
\end{split}
\end{equation} The last line confirms that the solution is indeed self-dual. The second line tells us that $Q_{-1} = 0 = Q_{1}$. From the first line, we see that, from the point of view of an observer near the horizon, the charges are not conserved ($Q_{-2} \neq 0$). This might appear surprising, as the Kerr-Taub NUT solutions is a stationary solution, and suggests that $Q_{-2}$ contains more information than just the radiation. Notice that the non-conservation of the charges in the leading phase space of self-dual Kerr-Taub NUT was already observed in \cite{Giribet:2025ihk} and the interpretation of the flux in that context remains an open question. Furthermore, we also see from the first line of \eqref{sdWeyl} that $Q_{0}, Q_2, Q_3 \neq 0$, which consitute natural observable quantities near a self-dual Kleinian Taub-NUT black hole horizon.

\section{Discussion}
\label{sec:Discussion}

In this work, we have discussed the solution space of general relativity around a null hypersurface at finite distance, and discussed the characteristic initial value problem in both metric and NP formalism. We have then exploited the power of the GHP formalism and introduced derivative operators to rewrite the Bianchi identities in a manifestly Weyl-covariant way. Furthermore, we have idenfied the Ashtekar-Steubel symplectic structure with the subleading symplectic structure at the horizon. This allowed us to identify the celestial $Lw_{1+\infty}$ symmetries at the horizon, originally found at null infinity. More precisely, we have constructed the canonical generators of the $Lw_{1+\infty}$ symmetries acting on the subleading phase space at the horizon.

This work opens new avenues that would be interesting to explore:
\begin{itemize}
    \item Symmetries of black hole horizons are referred to as soft hairs, and have been suggested to account for the microstate counting of black hole entropy. Famous examples where this counting has been achieved include BTZ black holes in AdS$_3$ \cite{Strominger:1997eq}, extremal black holes \cite{Guica:2008mu}, and has been conjectured to extend to non-extremal Kerr black holes using hidden symmetries \cite{Castro:2010fd}. It would be interesting to see whether these $Lw_{1+\infty}$ symmetries provide any further hints towards this direction. 

    \item The matching between null infinity and null hypersurface at finite distance has recently generated a lot of discussions \cite{Ashtekar:2024mme, Ashtekar:2024bpi, Riello:2024uvs,Ashtekar:2024stm,Riello:2024uvs,Bhambure:2024ftz,Ciambelli:2025mex}. Most of these works focus on the identification with the leading phase space at the horizon. We believe that our Weyl-covariant set-up provides further clarifications on that matter, and in particular of the fact that the Ashtekar-Steubel structure should be matched on the subleading symplectic structure at the horizon, instead of the leading one.  

\item So far, all the discussions about the {$Lw_{1+\infty}$} symmetries from a phase space perspective have been done using self-duality conditions and/or truncations of the equations of motion \cite{Freidel:2021ytz, Geiller:2024bgf, Kmec:2024nmu, Cresto:2024mne}. As we have discussed in the text, these conditions are even more crucial at finite distance, as the deviation between full and self-dual gravity appears at leading order. Despite these technical requirements, some of the structures are expected to survive beyond self-duality. Indeed, at infinity, patterns of these symmetries have been identified in the multipole expansion of the metric at null infinity \cite{Compere:2022zdz}, as well as in gravitational wave memory effect observables \cite{Grant:2021hga}. It would be interesting to repeat these discussions at finite distance and extract interesting observables beyond the self-dual sector.

\item This work provides further tools to include black holes in the framework of Carrollian and celestial holography, see \cite{Crawley:2023brz} for preliminary results in the self-dual sector. It would be interesting to map the holographic Carrollian/celestial CFT to the black hole horizon, and see the precise role of the subleading phase space in this context. It would also be interesting to understand the precise interplay between the charge aspects constructed here and the Carrollian momenta recently discussed in \cite{Fiorucci:2025twa,Hartong:2025jpp} and the role that the higher-spin charges have to play in this context.

\item The subleading phase space and charges have been discussed at null infinity in \cite{Godazgar:2018vmm,Godazgar:2018dvh,Godazgar:2020gqd,Godazgar:2020kqd} using covariant phase space methods. Remarkably, the Newman-Penrose conserved quantities \cite{Newman:1968uj} have been re-interpreted as arising from these subleading charges. Interestingly, similar conserved quantities exist for extremal black holes and are called the Aretakis charges. The correspondence between the Newman-Penrose and Aretakis charges has been discussed in \cite{Godazgar:2017igz} using a similar conformal rescaling that the one discussed here for general spacetimes. It would be interesting to understand whether the Aretakis charges are also captured by our subleading charges at the horizon.

\item It was recently shown for Yang-Mills theory that the higher spin symmetries, forming the $S$-algebra, can be interpreted as overleading gauge transformations \cite{Nagy:2022xxs,Nagy:2024dme,Nagy:2024jua}. This requires extending the standard radiative phase space with overleading Goldstone modes via a St\"uckelberg procedure. It would be interesting to understand if this procedure can be repeated for gravity at the horizon and understand the interplay with the subleading Ashtekar-Streubel phase space derived in this paper. 

\item The celestial $Lw_{1+\infty}$ symmetries possess a natural interpretation on twistor space as residual gauge transformations. It would be interesting to derive the surface charge expressions obtained here from first principles using similar method as in \cite{Kmec:2024nmu}. This would require defining an analgoue of asymptotic twistor space adapted to black hole horizons. We leave this interesting question for future endeavours.

\end{itemize}

%%%%%%%%%%%%%%%%%%%%%%%%%%%%%%%%%%%%%%%%%%%%%%%%%%%%%%%%%%%%%%%%%%%%%
\section*{Acknowledgments}
%%%%%%%%%%%%%%%%%%%%%%%%%%%%%%%%%%%%%%%%%%%%%%%%%%%%%%%%%%%%%%%%%%%%%
It is our pleasure to thank Luca Ciambelli, Nicolas Cresto, Adrien Fiorucci, Laurent Freidel, Gaston Giribet, Marc Geiller, Hernan Gonzalez, Daniel Grumiller, Adam Kmec, Robert Myers, Silvia Nagy, Marios Petropoulos,  Giorgio Pizzolo, Lionel Mason, Atul Sharma, and Simone Speziale for useful discussions.

RR is supported by the European Union’s Horizon Europe research and innovation programme under the Marie Skłodowska-Curie grant agreement No. 101104845 (UniFlatHolo), hosted at Harvard University and École Polytechnique. Research at Perimeter Institute is supported in part by the Government of Canada through the Department of Innovation, Science and Economic Development Canada and by the Province of Ontario through the Ministry of Colleges and Universities. The work of CZ was partially supported by research funds from the Solvay Family.

\appendix

\section{Kerr solution}\label{sec:Kerr}
\paragraph{Metric formalism}
The Kerr black hole in Gaussian null coordinates \eqref{metricGNC} can be found in \cite{Booth:2012xm}. The surface gravity $\kappa$ reads as
\begin{equation} 
\kappa = \frac{r_+ - r_-}{2(r_+^2 + a^2)},
\end{equation} 
where
\begin{equation}
r_\pm = M \pm \sqrt{M^2 - a^2}
\end{equation}
is the radius of the outer/inner event horizon, $ a = \frac{J}{M}$ with $J$ the angular momentum of the black hole and $M$ its mass.
Then we have explicitly
\begin{subequations}
\begin{align}
    q_{AB}&=(r_+^2+a^2\cos(\theta)^2)\extd\theta^2+\frac{\left(a^2+r_+^2\right)^2 \sin^2(\theta)}{a^2 \cos^2(\theta)+r_+^2}  \extd\phi^2 , \\
P^A&=\frac{a r_+^2 \left(a^2+3 r_+^2\right)+\left(a^3 r_+^2-a^5\right) \cos^2(\theta)}{\left(a^2+r_+^2\right)^2 \left(a^2 r_+ \cos^2(\theta)+r_+^3\right)}\partial_\phi+
\frac{a^2 \sin (2 \theta)}{\left(a^2 \cos^2(\theta)+r_+^2\right)^2}\partial_\theta , \\
\chi&=\frac{a^4 (\cos(2\theta)-1)-a^2 r_+^2 (\cos(2\theta)+7)-8 r_+^4}{r_+ \left(a^2+r_+^2\right) \left(a^2 (\cos(2\theta)+1)+2 r_+^2\right)} , \\
\chi_{\phi\theta}&=\frac{2 a^3 \left(a^2+r_+^2\right) \sin^3(\theta) \cos(\theta)}{\left(a^2 \cos^2(\theta)+r_+^2\right)^2} , \\
\chi_{\theta\theta}&=\frac{a^2 \sin^2(\theta) \left(a^4+a^2 \left(a^2-r_+^2\right) \cos(2\theta)-7 a^2 r_+^2-10 r_+^4\right)}{2 r_+ \left(a^2+r_+^2\right) \left(a^2 \cos(2\theta)+a^2+2 r_+^2\right)} . 
\end{align}
\end{subequations}
For Schwarzschild black hole, we set $a = 0$ so that $P^A=0$, $\chi_{AB}=0$ and $\chi=-\frac{4}{r_+}$. 

\paragraph{NP formalism}
\label{sec:KerrNP}
Using the dictionary between metric and NP formalism in Equation \eqref{dictionary Weyl scalars}, we deduce the expression of the Weyl scalars at the horizon of a  Kerr-AdS black hole:
\begin{subequations}
\begin{align}
\Psi_4^0&=\Psi_3^0=0\\     \Psi_2^0 &=-\frac{\Lambda }{6}+\frac{\left(a^2+r_+^2\right) }{2 r_+ \left(a^2 \cos^2(\theta)+r_+^2\right)^3} \left(r_+^3-3 a^2 r_+ \cos^2(\theta)+i \,a\cos(\theta)\left(a^2 \cos ^2(\theta)-3  r_+^2 \right)\right) , \\
%%%%%%%%%%%%%%%%%%%%%%%%%%%%%%%
\nonumber
\Psi_1^0&=\frac{a \sin (\theta)} {r_+ \left(a^2 \cos(2\theta)+a^2+2 r_+^2\right)^{7/2}}
\Big(a \left(9 a^4-15 a^2 r_+^2-20 r_+^4\right) \cos(\theta)+a^3 \left(3 a^2+7 r_+^2\right) \cos (3 \theta)  , \\
&+i
\left(2 a^4 r_+ \cos (4 \theta)-2  a^2 r_+ \left(7 a^2+11 r_+^2\right) \cos(2\theta)-2 r_+ \left(8 a^4+9 a^2 r_+^2+6 r_+^4\right)\right) \Big) . 
\end{align}
\end{subequations}

\section{Relevant equations of motion}
\label{sec:Relevant equations}

In this section, we write down useful equations in the NP formalism for which we introduce the notations
\begin{equation}
    \partial_0=m^A_0\partial_A\,,\qquad \bar\partial_0=\bar m^A_0\partial_A . 
\end{equation}
\paragraph{Metric equations}
\begin{subequations}
    \begin{align}\label{evolmA}
(-\partial_v+\gamma_0-\bar\gamma_0-\mu_0)m_0^A &=\bar\lambda_0 \bar m_0^A\,,\qquad 
%%%%%%%%%%%%%%%%%%%%%%%%%
(-\partial_v+\gamma_0-\bar\gamma_0+\mu_0)m^0_A=-\bar\lambda_0 \bar m^0_A , \\\label{evolmbA}
%%%%%%%%%%%%%%%%%%%%%%%%%
(-\partial_v-\gamma_0+\bar\gamma_0-\mu_0)\bar m_0^A &=\lambda_0 m_0^A\,,\qquad 
%%%%%%%%%%%%%%%%%%%%%%%%%
(-\partial_v-\gamma_0+\bar\gamma_0+\mu_0)\bar m^0_A=-\lambda_0 m^0_A . 
\end{align}
\end{subequations}

\paragraph{Evolution spin-coefficient equations}
We first have the equations sourced by the Weyl scalars. 
\begin{subequations}
\begin{align}\label{evolambda}
 \Psi^0_4&   =\left(\partial_v+2\mu_0+3\gamma_0-\bar\gamma_0\right)\lambda_0 , \\ \label{NPDamour2}
  \Psi^0_3&   =\left( \partial_v +\mu_0+\gamma_0-\bar\gamma_0\right)\alpha_0+\bar \partial_0\gamma_0+\lambda_0(\tau_0+\beta_0) , \\
\Psi^0_2&   =\left( \partial_v +\mu_0-\gamma_0-\bar\gamma_0\right)\rho_0+\bar \partial_0\tau_0+\lambda_0\sigma_0+2\tau_0\alpha_0+\frac\Lambda3 . 
\end{align}
\end{subequations}

The other evolution equations are  
\begin{subequations}
\begin{align}\label{NPDamour1}
&(\partial_v-\gamma_0+\bar\gamma_0+\mu_0)\beta_0=-\partial_0\gamma_0-\bar\lambda_0\alpha_0-\mu_0\tau_0 , \\\label{NPRaych}
&(\partial_v+\gamma_0+\bar\gamma_0+\mu_0)\mu_0+\lambda_0\bar\lambda_0 =0, \\\label{NPeqchi}
&(\partial_v-3\gamma_0+\bar\gamma_0+\mu_0)\sigma_0+(\partial_0+2\beta_0)\tau_0+\rho_0\bar\lambda_0=0. 
\end{align}
\end{subequations}
By combining different equations of motion we obtain 
\begin{align}\label{evoltau0}
&(\partial_v-\gamma_0+\bar\gamma_0+3\mu_0)\tau_0=\partial_0(\mu_0-\bar\gamma_0-\gamma_0)+(-\bar\partial_0-3\alpha_0+\bar\beta_0)\bar\lambda_0 . 
\end{align}

\paragraph{Bianchi identities - evolution equations}
\begin{subequations} 
\begin{align} 
\left(n^\mu\partial_\mu-2\gamma-4\mu\right)\Psi_3&=\left({m}^\mu\partial_\mu+1 \tau-4\beta\right)\Psi_4-3\nu\Psi_2.\label{NP3}\\
%%%%%%%%%%%%%%%%%%%%%%%%%%%%%%%%%%%%%
\left(n^\mu\partial_\mu + 0\gamma-3\mu\right)\Psi_2&=\left({m}^\mu\partial_\mu+2\tau-2\beta\right)\Psi_3-\sigma\Psi_4-2\nu\Psi_1,\label{NP2}\\
%%%%%%%%%%%%%%%%%%%%%%%%%%%%%%%%%%%%%
\left(n^\mu\partial_\mu+2\gamma-2\mu\right)\Psi_1&=\left({m}^\mu\partial_\mu+3\tau+0\beta\right)\Psi_2-2\sigma\Psi_3-\nu\Psi_0,\label{NP1}\\
%%%%%%%%%%%%%%%%%%%%%%%%%%%%%%%%%%%%%
\left(n^\mu\partial_\mu+4\gamma-1\mu\right)\Psi_0&=\left({m}^\mu\partial_\mu+4\tau+2\beta\right)\Psi_1-3\sigma\Psi_2 . \label{NP0}
%%%%%%%%%%%%%%%%%%%%%%%%%%%%%%%%%%%%%
\end{align}
\end{subequations}
At leading order we have
\begin{subequations}
\label{eq:Bianchi evolution}
\begin{align}\label{evolQ-1}
\left(-\partial_v-2\gamma_0-4\mu_0\right)\Psi^0_3&=\left(\partial_0-4\beta_0+1 \tau_0\right)\Psi^0_4.\\
%%%%%%%%%%%%%%%%%%%%%%%%%%%%%%%%%%%%%
\left(-\partial_v- 0\gamma_0-3\mu_0\right)\Psi^0_2&=\left(\partial_0-2\beta_0+2\tau_0\right)\Psi^0_3-\sigma_0\Psi^0_4,\\
%%%%%%%%%%%%%%%%%%%%%%%%%%%%%%%%%%%%%
\left(-\partial_v+2\gamma_0-2\mu_0\right)\Psi^0_1&=\left(\partial_0+0\beta_0+3\tau_0\right)\Psi^0_2-2\sigma_0\Psi^0_3,\\
%%%%%%%%%%%%%%%%%%%%%%%%%%%%%%%%%%%%%
\left(-\partial_v+4\gamma_0-1\mu_0\right)\Psi^0_0&=\left(\partial_0+2\beta_0+4\tau_0\right)\Psi^0_1-3\sigma_0\Psi^0_2 . 
%%%%%%%%%%%%%%%%%%%%%%%%%%%%%%%%%%%%%
\end{align}
\end{subequations}

\paragraph{Bianchi identities - radial equations} 
\begin{subequations}\label{Bianchiradial}
\begin{align}
\left(\ell^\mu\partial_\mu+4\rho\right)\Psi_1&=\left(\bar{m}^\mu\partial_\mu+4\alpha\right)\Psi_0,\label{NPB0}\\
%%%%%%%%%%%%%%%%%%%%%%%%%%%%%%%%%%%%%
\left(\ell^\mu\partial_\mu+3\rho\right)\Psi_2&=\left(\bar{m}^\mu\partial_\mu+2\alpha\right)\Psi_1+1\lambda\Psi_0,\label{NPB1}\\
%%%%%%%%%%%%%%%%%%%%%%%%%%%%%%%%%%%%%
\left(\ell^\mu\partial_\mu + 2\rho\right)\Psi_3&=\left(\bar{m}^\mu\partial_\mu+0\alpha\right)\Psi_2+2\lambda\Psi_1,\label{NPB2}\\
%%%%%%%%%%%%%%%%%%%%%%%%%%%%%%%%%%%%%
\left(\ell^\mu\partial_\mu+1\rho\right)\Psi_4&=\left(\bar{m}^\mu\partial_\mu-2 \alpha\right)\Psi_3+3\lambda\Psi_2.\label{NPB3}
\end{align}
\end{subequations}

\section{Diffeomorphism interpretation}
\label{sec:Diffeomorphism interpretation}
In this Appendix, we relate the spin $0$ and spin $1$ symmetries with the residual gauge transformations derived in Section \eqref{app:charges}.

We introduce
\begin{equation}
    \pounds_{(\mathcal Y ,\bar{\mathcal Y })} =\mathcal Y \bar\eth + \bar{\mathcal Y }\eth -\frac s2 (\eth + \bar\eth)(\mathcal Y -\bar{\mathcal Y })-(s+q)\left( \bar{\mathcal Y } \,\tau_0+\mathcal Y \,\bar\tau_0\right)
\end{equation}
where $ m_A^0Y^A=\mathcal Y $ of weight $\mathcal Y:\{1,-1\}$,  $\bar  m_A^0Y^A=\bar{\mathcal Y }$  of weight $\bar{\mathcal Y }:\{-1,1\}$ and $s=\frac{(p-q)}2$. We obtain the following transformation laws from the metric derivation \eqref{residualsymm}
\begin{equation}
    \delta_\xi \Psi_n^0= \left( f\partial_v +   \pounds_{(\mathcal Y ,\bar{\mathcal Y })}-s\, \partial_vf \right)\Psi_n^0 - (4-n) \eth f \, \Psi_{n+1}^0
\end{equation}
and
\begin{align}
\delta_\xi\lambda_0 & =(f\partial_v+\pounds_{(\mathcal Y ,\bar{\mathcal Y })} + \partial_vf)\lambda_0 \,,\quad 
%%%%%%%%%%%%%%%%%%%%%%%%%%%%
\delta_\xi\sigma_0  =(f\partial_v+\pounds_{(\mathcal Y ,\bar{\mathcal Y })}-\partial_vf)\sigma_0 -\eth^2f%-2\tau_0\eth f
%%%%%%%%%%%%%%%%%%%%%%%%%%%%
\end{align}
where we assume $f:\{0,0\}$ and we set $\tau_0=0$. 
The term $\partial_vf$ in the transformation laws is related to the Weyl weight, and for that reason we view $f$ and $\partial_vf$ as two independent variables. We identify $\partial_vf \,\eta=-3\kappa\,f\,\eta+\frac{s_\eta}{2}(\eth + \bar\eth)(\mathcal Y -\bar{\mathcal Y })\eta$.  
These transformations reproduce the action of $T_0$ \eqref{delta0sigma}, \eqref{delta0lambda} and $T_1$  \eqref{delta1sigma}, \eqref{delta1lambda} provided that 
\begin{align}
    f&=-(T_0 +U\,\eth T_1) \,, \quad \mathcal Y =0\,,  \quad  \bar{\mathcal Y } =2T_1 \quad  \text{with } \bar\eth T_1=-\frac14 \eth T_1 . 
\end{align}
 We leave to future work a more detailed symmetry analysis of the phase space.

\section{Radial expansion}
\label{sec:Radial expansion}

In this Appendix, following a similar procedure than the one at null infinity \cite{Freidel:2021ytz}, we identify higher spin charge aspects $Q_s$ by looking at the subleading orders in the radial expansion of the Weyl tensor. Solving the radial Bianchi identities \eqref{Bianchiradial},
we find
\begin{subequations}
\begin{align}
\Psi_4 &= \Psi^0_4 + r (\bar{\eth} \Psi^0_3 + 3 \lambda_0 \Psi^0_2 - \rho_0\Psi^0_4) + \mathcal{O}(r^2) , \\
  \Psi_3 &= \Psi^0_3 + r (\bar{\eth} \Psi^0_2 + 2 \lambda_0 \Psi^0_1 - 2\rho_0\Psi^0_3) + \mathcal{O}(r^2) , \\
  \Psi_2 &= \Psi^0_2 + r (\bar{\eth} \Psi^0_1 + 1\lambda_0 \Psi^0_0 - 3\rho_0\Psi^0_2) + \mathcal{O}(r^2) , \\
  \Psi_1 &= \Psi^0_1 + r (\bar{\eth} \Psi^0_0+ {0}\phantom{ \lambda_0 \Psi^0_0} - 4\rho_0\Psi^0_1) + \mathcal{O}(r^2) . 
\end{align}
\end{subequations}
Recalling \eqref{Weyl scalars}, a natural guess to obtain higher spin momenta would be to consider
\begin{equation}
    \Psi_0 = \Psi^0_0 + r (\bar{\eth} {Q}_3 - \lambda_0 {Q}_4 - 5\rho_0 \Psi^0_0) + \mathcal{O}(r^2) . 
\label{ansatz higher spin}
\end{equation} However, because of the presence of the $\lambda_0$ term in this equation, the situation is drastically different compared to infinity. Indeed, defining $Q_3$ form $\Psi^1_0$ would require knowing $Q_4$, and so on. An alternative way to interpret this equation is the following: suppose $Q_3$ is given once for all. Then $Q_4$ can be extracted from $\Psi^1_0$ and is determined by $Q_2 = \Psi^0_0$ and $Q_3$. This procedure can in principle be iterated to obtain all the higher-spin aspects $Q_s$ from the radial expansion. We leave for future work to check that a (consistent truncation of) the time evolution equations \eqref{Bianchi evolution equations} are compatible with \eqref{Bianchi evolution equations} for the higher-spin charge aspects defined through \eqref{ansatz higher spin}.

\section*{References}
\bibliographystyle{style}
\renewcommand\refname{\vskip -1.3cm}%removes the title of the references
\bibliography{Biblio}

\end{document}